\documentclass[graybox]{svmult}

\usepackage{float}         
\usepackage{placeins}      

\usepackage{type1cm}        
\usepackage{makeidx}         
\usepackage{graphicx}        
\usepackage{multicol}        
\usepackage[bottom]{footmisc}

\usepackage{amsfonts} 
\usepackage{dsfont}

\usepackage{algorithm}
\usepackage{algpseudocode}

\usepackage{subfig}
\usepackage{nth}

\usepackage{glossaries}
\newacronym{5G}{5G}{Fifth Generation}
\newacronym{6G}{6G}{Sixth Generation}
\newacronym{AE}{AE}{Auto-Encoder}
\newacronym{AoA}{AoA}{Angle of Arrival}
\newacronym{ADC}{ADC}{Analog-to-Digital Convertor}
\newacronym{AWGN}{AWGN}{Additive White Gaussian Noise}
\newacronym{BER}{BER}{bit error rate}
\newacronym{BS}{BS}{Base Station}
\newacronym{CE}{CE}{Cross Entropy}
\newacronym{CNN}{CNN}{Convolutional Neural Network}
\newacronym{CS}{CS}{Compressed Sensing}
\newacronym{CSI}{CSI}{Channel State Information}
\newacronym{DAC}{DAC}{Digital-to-Analog Convertor}
\newacronym{DC}{DC}{Direct Current}
\newacronym{D2D}{D2D}{Device-to-Device}
\newacronym{DeepSC}{DeepSC}{Deep Semantic Communications}
\newacronym{DMA}{DMA}{Dynamic Metasurface Antenna}
\newacronym{DNN}{DNN}{Deep Neural Network} 
\newacronym{DTFT}{DTFT}{discrete-time Fourier transform}
\newacronym{E2E}{e2e}{End-to-End}
\newacronym{EI}{EI}{Edge Inference}
\newacronym{EM}{EM}{ElectroMagnetic}
\newacronym{FFNN}{FFNN}{Feed-Forward Neural Network}
\newacronym{GOC}{GOC}{Goal-Oriented Communications}
\newacronym{HRIS}{HRIS}{Hybrid Reconfigurable Intelligent Surface}
\newacronym{iid}{i.i.d.}{independent and identically distributed}
\newacronym{IoT}{IoT}{Internet of Things}
\newacronym{ISAC}{ISAC}{Integrated Sensing and Communications}
\newacronym{ISI}{ISI}{intersymbol interference}
\newacronym{JSCC}{JSCC}{Joint Source Channel Coding}
\newacronym{LTI}{LTI}{linear time-invariant}
\newacronym{LoS}{LoS}{Line-of-Sight}
\newacronym{MAP}{MAP}{maximum a-posteriori probability}
\newacronym{MEC}{MEC}{Multi-access Edge Computing}
\newacronym{MI}{MI}{Mutual Information}
\newacronym{MIMO}{MIMO}{Multiple-Input Multiple-Output}
\newacronym{MINN}{MINN}{Metasurfaces-Integrated Neural Network}
\newacronym{MISO}{MISO}{Multiple Input Single Output}
\newacronym{ML}{ML}{Machine Learning}
\newacronym{MLP}{MLP}{Multi Layer Perceptron}
\newacronym{mmWave}{mmWave}{millimeter wave}
\newacronym{mMTC}{mMTC}{massive Machine-Type Communications}
\newacronym{MS}{MS}{Meta-Surface} 
\newacronym{MSE}{MSE}{Mean Squared Error}
\newacronym{NLoS}{NLoS}{Non Line-of-Sight}
\newacronym{OAC}{OAC}{Over-the-Air Computation}
\newacronym{OTA}{OTA}{Over-the-Air}
\newacronym{PDF}{PDF}{probability density function}
\newacronym{PHY}{PHY}{PHYsical}
\newacronym{PIN}{PIN}{Positive-Intrinsic-Negative}
\newacronym{PSK}{PSK}{Phase Shift Keying}
\newacronym{PSD}{PSD}{power spectral density}
\newacronym{ReLU}{ReLU}{Rectified Linear Unit}
\newacronym{RF}{RF}{Radio-Frequency}
\newacronym{RICS}{RICS}{Reconfigurable Intelligent Computational Surface}
\newacronym{RIS}{RIS}{Reconfigurable Intelligent Surface}
\newacronym{RS}{RS}{Reed-Solomon}
\newacronym{RV}{RV}{random variable}
\newacronym{RX}{RX}{Receiver}
\newacronym{SER}{SER}{symbol error rate}
\newacronym{SIM}{SIM}{Stacked Intelligent Metasurfaces}
\newacronym{SINR}{SINR}{signal-to-interference-and-noise ratio}
\newacronym{SISO}{SISO}{Single Input Single Output}
\newacronym{SNR}{SNR}{Signal-to-Noise Ratio}
\newacronym{SotA}{SotA}{State of the Art}
\newacronym{STAR}{STAR}{Simultaneously Transmitting And Reflecting}
\newacronym{SGD}{SGD}{Stochastic Gradient Descent}
\newacronym{SVD}{SVD}{Singular Value Decomposition}
\newacronym{TDD}{TDD}{time division duplexing}
\newacronym{THz}{THz}{TerraHertz}
\newacronym{TX}{TX}{Transmitter}
\newacronym{UE}{UE}{User Equipment}
\newacronym{UT}{UT}{User Terminal}
\newacronym{VAE}{VAE}{Variational Auto-Encoder}
\newacronym{WMMSE}{WMMSE}{Weighted Minimum Mean Square Error}
\newacronym{WSS}{WSS}{wide-sense stationary}
\newacronym{GPU}{GPU}{Graphics Processing Unit}
\newacronym{D2NN}{D\textsuperscript{2}NN}{Deep Diffractive Neural Network}
\newacronym{RFC}{RFC}{Radio-Frequency Chain}
\newacronym{SLFN}{SLFN}{Single hidden Layer Feedforward Network}
\newacronym{ELM}{ELM}{Extreme Learning Machine}

\newcommand{\myvec}[1]{{\boldsymbol{#1}}}

\usepackage{newtxtext}       %
\usepackage{newtxmath}       

\usepackage{xcolor}

\makeindex             

\begin{document}

\title*{Over-the-Air Goal-Oriented Communications}
\author{
Kyriakos Stylianopoulos \\
Paolo Di Lorenzo \\
George C. Alexandropoulos}
\authorrunning{K. Stylianopoulos, P. Di Lorenzo, and G. C. Alexandropoulos}

\institute{
Kyriakos Stylianopoulos \at Department of Informatics and Telecommunications, National and Kapodistrian University of Athens,  Greece, \email{kstylianop@di.uoa.gr} \\
\\
Paolo Di Lorenzo \at National Inter-University Consortium for Telecommunications (CNIT), Parma, Italy and DIET Department, Sapienza University of Rome, Rome, Italy,
\email{paolo.dilorenzo@uniroma1.it} \\
\\
George C. Alexandropoulos \at Department of Informatics and Telecommunications, National and Kapodistrian University of Athens,  Greece, \email{alexandg@di.uoa.gr} \\
}
%
%
\maketitle

Goal-oriented communications offer an attractive alternative to the Shannon-based communication paradigm, where the data is never reconstructed at the Receiver (RX) side.
Rather, focusing on the case of edge inference, the Transmitter (TX) and the RX cooperate to exchange features of the input data that will be used to predict an unseen attribute of them, leveraging information from collected data sets.
This chapter demonstrates that the wireless channel can be used to perform computations over the data, when equipped with programmable metasurfaces.
The end-to-end system of the TX, RX, and MS-based channel is treated as a single deep neural network which is trained through backpropagation to perform inference on unseen data.
Using Stacked Intelligent Metasurfaces (SIM), it is shown that this Metasurfaces-Integrated Neural Network (MINN) can achieve performance comparable to fully digital neural networks under various system parameters and data sets. By offloading computations onto the channel itself, important benefits may be achieved in terms of energy consumption, arising from reduced computations at the transceivers and smaller transmission power required for successful inference.

\section{Introduction}
The rapid proliferation of connected devices, driven by the \gls{IoT} paradigm, has made \gls{mMTC} a cornerstone of \gls{5G} networks~\cite{8624514}. Looking toward \gls{6G} systems and beyond, the focus is shifting towards realizing ultra low latency and energy-efficient \gls{D2D} wireless links. This necessitates a fundamental reconsideration of the \gls{PHY} layer design, calling for innovations in power- and cost-efficient hardware combined with significant advancements in information processing algorithms~\cite{6g_keydrivers_2019}.
To manage the enormous volume of \gls{IoT}-generated data, which includes signals used for positioning and sensing, processing at the network edge is becoming essential. \gls{6G} is thus envisioned to embrace a cross-layer design, blurring the traditional boundaries between the user plane and the \gls{PHY} layer. This data-driven trend promotes \gls{EI}~\cite{9606720} and \gls{GOC}~\cite{LWM23}.

In this new paradigm, the \gls{TX} sends data not for perfect reconstruction at the \gls{RX}, but to enable the \gls{RX} to extract information necessary for a specific network task. \gls{EI} is a specialized form of this approach, focusing on enabling the \gls{RX} to infer a target feature from the transmitted symbol using patterns learned from past input-target examples.
This offers dual benefits: reduced computational complexity at the \gls{RX} (by avoiding full data reconstruction) and efficient use of communication resources (by encoding only task-relevant information).
\gls{ML} tools~\cite{Alexandropoulos2022Pervasive}, particularly \glspl{DNN}~\cite{alexandropoulos2020phaselearning_all, Multi_Hop, Stamatelis_NE, Stylianopoulos_DRL_1bit}, are gaining traction for \gls{IoT} applications at the network edge in \gls{6G}~\cite{6G-AI-china, Stylianopoulos_Autoregressive}.
They offer a data-driven alternative to traditional model-based approaches, which often rely on unrealistic assumptions. \gls{ML} excels at identifying and exploiting complex patterns that accurately reflect the deployment environment.
Moreover, the heavy computational load associated with \gls{ML} is typically confined to the offline training phase, enabling low latency computations during deployment.
However, a major obstacle for \gls{EI} is the hardware complexity and resulting power consumption. The efficient execution of \gls{DNN} computations relies heavily on parallel processing units, which substantially increase energy demands at the edge.
A truly transformative concept proposes that computational tasks need not be restricted solely to digital transceivers.
By developing \gls{GOC} based on smart wireless environments~\cite{RIS_challenges_all,di2019smart}, the wireless channel can be transformed from a passive medium into an active computational entity~\cite{YCX23_Wave_Computing,ASD21}.

This is made possible through the use of (programmable) \glspl{MS}.
Those near-\gls{EM}-passive structures, deployed as \glspl{RIS} for controlled reflection~\cite{huang2019reconfigurable,BAL24_RIS_review}, or as \gls{SIM}~\cite{AXN23, ranasinghe2025_parameterized_SIM, Huang19_SPAWC} at the \gls{TX}/\gls{RX} for extremely large apertures~\cite{huang2019holographic_all}, offer controllable wave transformations~\cite{OTA_RISs}.
Traditionally, such \gls{MS} technologies have been developed for low complexity and energy efficient signal strength amplification in communication and sensing applications~\cite{JAS22_LowToZero, Moustakas_LargeSystemAnalysis, Stylianopoulos_Asymptotic_1bit, Gavriilidis25_Asilomar, Gavriilidis_NF_Beam_Tracking}.
By shaping the propagating \gls{EM} waves \gls{OTA} using passive or analog operations, these systems can execute portions of feature extraction, compression, or filtering with almost zero energy consumption.
This process effectively offloads computational tasks from conventional, energy-hungry digital hardware into the \gls{PHY} layer itself, paving the way for sustainable and ultra-efficient \gls{ML}-enabled systems and \gls{EI} applications~\cite{YCX23_Wave_Computing}.

In this context, this chapter explores the joint exploitation of \glspl{MS} and conventional wireless effects to realize computations analogous to those performed by \glspl{DNN}. This approach allows the entire \gls{E2E} \gls{MS}-parametrized \gls{MIMO} system to be treated as a single \gls{DNN}, composed of digital, analog, and wave-domain layers of computation, offering a significant reduction in the complexity and power requirements for \gls{IoT} devices performing \gls{EI}.

\section{Preliminaries of Goal-Oriented Communications}\label{sec:goc-review}

\subsection{Probabilistic Inference}\label{sec:inference-theory}
The objective of an inference task is to determine a target attribute $\myvec{o}$ from a given input observation $\myvec{x}$, represented by the unknown mapping $\myvec{o} = l(\myvec{x})$. Since an analytical form for $l(\myvec{x})$ is intractable, this relationship is approximated using a dataset of input-target pairs, $\mathcal{D} \triangleq \{ (\myvec{x}_i, \myvec{o}_i) \}_{i=1}^{|\mathcal{D}|}$.
From a probabilistic perspective, solving the inference problem involves fitting the conditional \gls{PDF} $p(\myvec{o}|\myvec{x})$ to the data.
However, in most practical scenarios, only point estimates are required, reducing the task to predicting the most probable target value ($\hat{\myvec{o}}$) for a given observation $\myvec{x}$.

In the \gls{ML} paradigm, this approximation is handled by a parametrized model: $\hat{\myvec{o}} \triangleq f_{\myvec{w}}(\myvec{x})$. The parameter values $\myvec{w}$ are optimized by minimizing an amortized cost function, $J(\myvec{w})$, averaged over all training instances in $\mathcal{D}$:
\begin{equation}\label{eq:cost-function}
   J(\myvec{w}) \triangleq \frac{1}{|\mathcal{D}|} \sum_{i=1}^{|\mathcal{D}|} \mathfrak{J}(\myvec{o}_i, \hat{\myvec{o}}_i), \quad\text{where } \hat{\myvec{o}}_i = f_{\myvec{w}}(\myvec{x}_i). 
\end{equation}
The per-instance cost, $\mathfrak{J}(\myvec{o}_i, \hat{\myvec{o}}_i)$, quantifies the prediction error and often has a direct probabilistic interpretation~\cite{MacKay2003, deep_learning_book}:

\begin{itemize}
    \item \textbf{Classification:} For tasks where the input is assigned to one of $d_{\rm cl}$ classes (target $\myvec{o}$ is one-hot encoded, i.e., $\myvec{o}$ is a $d_{\rm cl}$-sized vector containing zeros everywhere except for the element and the index of the input class which contains one), the standard choice is the \gls{CE} loss function:
    \begin{equation}\label{eq:cross-entropy}
        \mathfrak{J}_{\rm CE}(\myvec{o}_i, \hat{\myvec{o}}_i) \triangleq - \sum_{j=1}^{d_{\rm cl}} [\myvec{o}_i]_j \log [\hat{\myvec{o}}_i]_j.
    \end{equation}
    Minimizing $\mathfrak{J}_{\rm CE}$ is mathematically equivalent to performing maximum likelihood estimation of the model parameters ($\myvec{w}$) under the assumption that the conditional \gls{PDF} $p(\myvec{o}|\myvec{x})$ follows a multivariate Bernoulli distribution.

    \item \textbf{Regression:} For continuous target predictions, where $\myvec{o}, \hat{\myvec{o}} \in \mathbb{R}^{d_{\rm out} \times 1}$, the common \gls{MSE} metric ($1/d_{\rm out} \left\|\myvec{o} - \hat{\myvec{o}}\right\|^2$) is typically used. The use of \gls{MSE} implies the assumption that the conditional \gls{PDF} $p(\myvec{o}|\myvec{x})$ is a Gaussian distribution.
\end{itemize}

\subsection{Artificial Neural Networks}\label{sec:NN-theory}
While a broad spectrum of parametrized function families is available for modeling $f_{\myvec{w}}(\cdot)$, the current cutting edge in inference relies heavily on \glspl{DNN}. These networks are favored due to their remarkable expressivity, a wide array of specialized architectural components tailored to various tasks, and the inherent parallelizability of their computations, which allows for real-time inference on modern hardware. Mathematically, a neural network is defined as a composition of $L$ layers:
\begin{equation}\label{eq:neuralnet}
f_{\myvec{w}}(\myvec{x}) \triangleq f^{L}_{\myvec{w}^L}\left( f^{L-1}_{\myvec{w}^{L-1}}\left(\ldots f^{1}_{\myvec{w}^1}( \myvec{x} ) \ldots \right)\right),
\end{equation}
where the $l$-th layer ($l=1,2,\ldots,L$) is governed by parameters $\myvec{w}_l$. The output of layer $l$, denoted by $\bar{\myvec{o}}^l$, serves as the input to the subsequent $(l+1)$-th layer, yielding the recursive definition $\bar{\myvec{o}}^l \triangleq f^{l}_{\myvec{w}^l}( \bar{\myvec{o}}^{(l-1)} )$. The network begins with the observation, $\bar{\myvec{o}}^0 = \myvec{x}$. For notational simplicity, the complete set of parameters is aggregated as $\myvec{w} \triangleq [\myvec{w}^{\top}_1, \myvec{w}^\top_2, \dots, \myvec{w}^\top_L]^\top$.

We will not go deeper into the specific details of individual layer functions here, acknowledging the extensive body of research dedicated to developing specialized layers that efficiently perform data-specific computations and extract high-level patterns from training sets \cite{deep_learning_book}. Nonetheless, a crucial requirement is that each layer function $f^l_{\myvec{w}_l}(\cdot)$, must incorporate a non-linear element. Historically, these nonlinearities were often discriminatory or sigmoidal~\cite{Cybenko89}. These properties are foundational to the \emph{universal approximation theorem}, which guarantees that artificial neural networks comprising at least two layers (and potentially infinite width) can approximate any arbitrary mapping $\myvec{o} = l(\myvec{x})$, thus achieving $f_{\myvec{w}}(\myvec{x})\cong l(\myvec{x})$~\cite{Cybenko89}.

Beyond these theoretical assurances, the practical task of determining the optimal parameter values $\myvec{w}$ for successful inference can be solved efficiently. This involves substituting the neural network expression from \eqref{eq:neuralnet} into the appropriate cost function (e.g., \eqref{eq:cross-entropy} for classification) and then into the amortized loss function \eqref{eq:cost-function}. This minimization problem is typically tackled using variants of the \gls{SGD} method. The core mechanism is the computation of gradients $\partial J(\myvec{w})/\partial\myvec{w}$ by leveraging the chain rule to propagate error signals backward through the network layers. This procedure defines the renowned \emph{backpropagation} algorithm~\cite{Rumelhart86, LeCun1989}, which is the central computational method underpinning deep learning.

\subsection{Edge Inference}\label{sec:EI-theory}
The emerging field of \gls{EI} involves the deployment and training of inference tasks across a wireless communication infrastructure. We consider an uplink scenario where a multi-antenna \gls{TX} observes the input $\myvec{x}$ and aims to communicate its estimate $\hat{\myvec{o}}$ of the target attribute $\myvec{o}$ to a remote \gls{RX}. Initially, this process appears achievable under two established communication paradigms:

\textbf{``Infer-then-transmit''}:
In this option, the \gls{TX} first computes the target estimate $\hat{\myvec{o}} = f_{\myvec{w}}(\myvec{x})$ locally. It then prepares the transmission signal $\myvec{s}$ by applying source coding (data compression) and channel coding (modulation and beamforming) to $\hat{\myvec{o}}$. These coding steps are necessary to ensure the system achieves a satisfactory communication rate, enabling the \gls{RX} to successfully reconstruct $\hat{\myvec{o}}$ via decoding and decompression.
Implementing these operations in high complexity modern wireless systems often requires dedicated optimization procedures and accurate \gls{CSI}, incurring extra computational costs.
While inference targets are typically of a much smaller dimension than the original observations, leading to small rate requirements, this approach burdens the \gls{TX} with the cost of executing \gls{DNN} computations locally. This hardware assumption is often overly optimistic for \gls{IoT} or other lightweight devices envisioned for \gls{EI} tasks, which are constrained by low complexity and minute power consumption.

\textbf{``Transmit-then-infer''}:
The converse approach bypasses local \gls{DNN}-based \gls{EI} at the \gls{TX}. Instead, the \gls{TX} performs only source and channel coding on the original observation $\myvec{x}$, which is then transmitted over the link. Subsequently, the \gls{RX} decodes the received signal to obtain the input data point and then feeds it into its local $f_{\myvec{w}}(\cdot)$ to perform inference. Although the \gls{RX} is typically assumed to possess sufficient power and hardware capacity to support a \gls{DNN} in uplink settings, transmitting the entire original observation $\myvec{x}$ often imposes high link budget demands that may be difficult to satisfy.
A flexible alternative to mitigate the trade-offs of the above options is possible by exploiting the sequential nature of the \gls{DNN} structure defined in~\eqref{eq:neuralnet}~\cite{dnn-splitting}. This leads to the following \gls{EI} paradigm:

\textbf{``Infer-while-transmitting'' (DNN splitting)}:
Intermediate representations, $\bar{\myvec{o}}^l$ for $l=1,2\ldots,L-1$, can be constructed with arbitrary dimensions. Architectural design often includes one or more small-sized \emph{bottleneck layers} (as seen in auto-encoders \cite{VAE} and U-Nets \cite{U_Net}), which effectively perform compression to preserve only the most relevant information. Leveraging this, one can partition the \gls{DNN} such that the first $L'<L$ layers reside at the \gls{TX}. The output $\bar{\myvec{o}}^{L'}$ is then transmitted over the network and passed sequentially to the remaining layers, from $(L'+1)$ up to $L$, at the \gls{RX}.

The latter paradigm is clearly the most flexible, allowing for dynamic balancing of the computational load and communication resources between the \gls{TX} and \gls{RX}.
The initial application of \glspl{DNN} in transceivers was studied under the \gls{JSCC} paradigm~\cite{DCH18, BKG19}. Unlike Shannon's separation theorem, which treats source and channel coding separately \cite{shannon}, \gls{JSCC} develops joint encoders and decoders that account for channel conditions to achieve superior reconstruction performance. However, \gls{JSCC}'s objective is limited strictly to data reconstruction.
Conversely, \gls{DeepSC} approaches~\cite{GQA23} aim to transmit the meaning of the data, quantified by cost functions defined by various \gls{GOC} objectives. For example, the \gls{DeepSC} architecture of the work in~\cite{DeepSC} uses separate source and channel sub-modules. While its channel encoder/decoder maximizes a difficult-to-evaluate \gls{MI} objective, this maximization negates the channel's effects rather than leveraging them for computation. Similarly, a separate source/channel coding approach for image retrieval~\cite{JGM21_Image_Retrieval} showed benefits but also treated the channel purely as a source of noise.
The concept of \gls{DNN} splitting for \gls{EI}~\cite{dnn-splitting, PBD22_bottleneck} has been investigated from an information bottleneck perspective to determine optimal network partitioning in uncontrollable wireless channels. This framework, which transmits intermediate feature representations, aligns well with the broader \gls{GOC} objective.

\subsection{Computational Considerations}\label{sec:computation-of-EI}
When implementing \gls{DNN} splitting over a wireless channel characterized by realistic effects (i.e., large- and small-scale fading, and \gls{AWGN}), the transmitted intermediate output $\bar{\myvec{o}}^{L'}$ from the $L'$-th \gls{DNN} layer will inevitably be distorted upon arrival at the \gls{RX}. Representing the channel state via an abstract random variable $\boldsymbol{\mathcal{H}}$, the training objective remains the minimization of the $\myvec{w}$-parametrized cost function $J(\myvec{w})$, but must now account for the stochastic nature of the wireless environment:
\begin{equation*}\label{eq:EI-objective}
\mathcal{OP}_{\rm EI}: \min_{\myvec{w}} \mathbb{E}_{\boldsymbol{\mathcal{H}}}[J(\myvec{w})],
\end{equation*}
where the value of the objective function is instantaneously affected by distortion induced by each channel realization.

Assuming sufficient wireless channel capacity, the standard paradigm of wireless communications suggests that optimizing both endpoints for source and channel encoding/decoding will effectively eliminate the distortion on the received $\bar{\myvec{o}}^{L'}$.
The resulting decoded output can then be directly fed into the network's $(L'+1)$-layer as if the channel were not present; the channel is effectively \emph{hidden} from the neural network's processing perspective.
This reconstruction-centric approach aligns with the traditional practices of both wireless communications and \gls{ML}, and it has successfully exhibited satisfactory results~\cite{DeepSC, HMC24_Mattias_RIS}.
However, optimizing the system primarily for signal reconstruction may result in unnecessary computational overheads due to the following considerations:
\begin{enumerate}
\item \gls{EI}'s primary objective is the calculation of an arbitrary function of the input, not the perfect reconstruction of intermediate variables. From this viewpoint, devoting resources to reconstructing intermediate variables is not always the most efficient path. \gls{GOC} can be viewed as a specialized instance of lossy compression between the unseen target $\myvec{o}$ and its estimation $\hat{\myvec{o}}$, where $\mathfrak{J}(\myvec{o}, \hat{\myvec{o}})$ acts as the distortion metric. Information-theoretic perspectives, such as those studying variations of the distortion-rate function \cite{SK22_Rate_Distorion}, indicate that the channel rate necessary to meet a desired error threshold for transmitting intermediate variables is typically less than the channel capacity required for reconstruction with arbitrarily small error probabilities.
\item From an engineering standpoint, perfect reconstruction may not be necessary, as the subsequent neural network layers are often inherently designed to tolerate noisy inputs, reflecting the stochastic nature of inference problems. Furthermore, deliberately injecting noise into layer activations during training \cite{Dropout, GMD16_Noisy_activations} and inference \cite{Variational_Dropout, MC_Dropout} is a recognized technique for enhancing model regularization and uncertainty estimation.
\item The wireless channel, which inherently acts as a stochastic function on the transmitted data, imposes its own computations. While this channel function is generally not controllable by the \gls{E2E} system, \gls{OAC} methods exploit the superimposition property of wireless signals to implement certain classes of computational functions directly on the propagation medium. Crucially, the controllability introduced by emerging \gls{MS} technologies has the potential to enable more sophisticated \gls{OAC}, effectively offloading computation away from the communication network endpoints.
Notice that, in this book, we differentiate \gls{OAC} from ``AirComp,'' as defined in~\cite{AirComp_Review}, which strictly addresses nomographic functions requiring predefined analytical functions at transceivers for computation, primarily applied in multiple access and federated learning. Our use of \gls{OAC} refers to any effects of the propagation environment on signals that are controllable and leveraged toward a specific computational goal.
\end{enumerate}

\section{Deep Diffractive Neural Networks}

\subsection{Basic Principles of Diffractive Computing}

\begin{figure}
    \centering
    \includegraphics[scale=0.8]{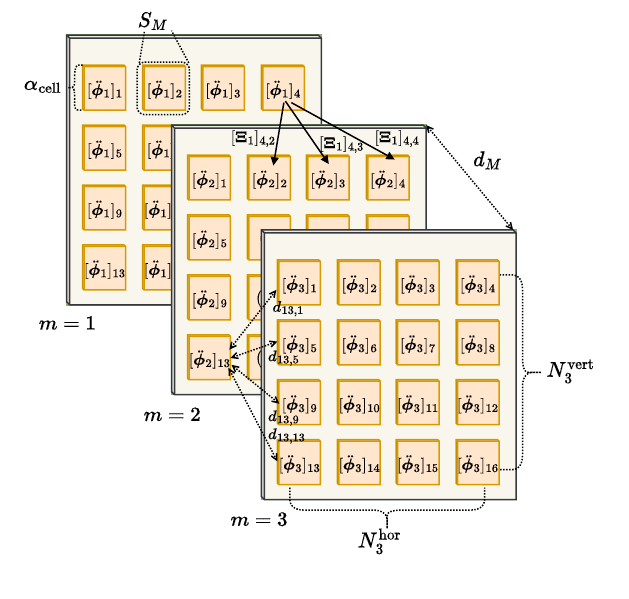}
    \caption{A \gls{SIM} of $M=3$ layers, each with $N_m=16$ elements, that is used as a \gls{D2NN} for wave-domain-based \gls{ML}. The controllable responses $[\ddot{\boldsymbol{\phi}}_m]_n$ ($m=1,\dots,M$ and $n=1,\dots,N$) are treated as trainable \gls{DNN} weights.}
    \label{fig:sim_notation}
\end{figure}

The primary technology enabling the wave domain implementation of \glspl{DNN} is the concept of \gls{SIM}~\cite{AXN23}.
A \gls{SIM} consists of densely arranged, thin layers of diffractive \glspl{MS}, where each \gls{MS} layer is composed of multiple unit elements offering tunable \gls{EM} responses.
The complete \gls{SIM} structure is typically housed within absorbing material, negating multipath effects.
By precisely controlling the \gls{EM} responses of the unit elements, specific linear operations can be executed on the signals propagating from the initial \gls{MS} layer.
Consider a \gls{SIM} device comprising $M$ layers of diffractive \glspl{MS}, each containing $N_m = N_m^{\rm vert} N_m^{\rm hor}$ elements arranged into $N_m^{\rm hor}$ columns and $N_m^{\rm vert}$ rows ($m=1,\dots,M$), as depicted in Fig.~\ref{fig:sim_notation}.
The element-to-element propagation between consecutive \gls{SIM} layers is governed by geometric optics due to their dense placement~\cite{AXN23, GJZ24_SIM_TOC, Stylianopoulos_GO}.
Given elements $n$ and $n'$ ($1 \leq n,n' \leq N$) with distance $d_{n,n'}$ and area $S_M$ from layers $m$  and $m-1$ ($2  \leq \! m \leq  M$), the propagation matrix $\mathbf{\Xi}_m \in  \mathbb{C}^{N \times N}$ is given via the Rayleigh-Sommerfeld diffraction equation~\cite{XYN18, AXN23}:
\begin{align}\label{eq:rayleigh-sommerfeld}
    [\mathbf{\Xi}_m]_{n,n'} &\triangleq \frac{d_M S_M}{d^2_{n,n'}} 
    \Big( \frac{1}{2\pi d_{n,n'}} - \frac{\jmath}{\lambda} \Big) 
    \exp\left({\jmath 2\pi d_{n,n'}}\right),
\end{align}
where $\lambda$ is the carrier frequency and $\jmath\triangleq\sqrt{-1}$ is the imaginary unit.
The responses of the unit elements of the $m$-th layer $\boldsymbol{v}_l$ are modeled as typical idealized unit-amplitude phase shifters, i.e., $[\ddot{\boldsymbol{\phi}}_m]_n \triangleq \exp(\jmath \xi_{m,n})$, where $\xi_{m,n}$ is the controllable phase shift.
It is important to highlight that, in deriving the exact form of~\eqref{eq:rayleigh-sommerfeld}, it is necessary to assume that the width of each \gls{MS} element is negligible to the layer-to-layer distance $d_M$, which provides a limit on how densely stacked the \gls{SIM} layers might be so that they are accurately modeled through this approach.
Considering that the width of each rectangular element is typically set to $\lambda/2$ (i.e., $S_M = \lambda^2/4$), a rule of thumb is to set $d_M \geq 5 \lambda$ to satisfy this assumption.
Contrarily, placing \gls{SIM} layers at uncharacteristically large distances, say $d_M \geq 10 \lambda$, induces large attenuation as it can be observed by the $d^3_{m,m'}$ factor at the denominator of~\eqref{eq:rayleigh-sommerfeld}, and may thus decrease the effectiveness of the overall \gls{SIM} device.

By letting $\ddot{\myvec{\Phi}}_m(t) \triangleq {\rm diag}(\ddot{\boldsymbol{\phi}}_m(t))$, the overall \gls{SIM} response matrix is mathematically expressed as \cite{RSR25}:
\begin{equation}\label{eq:sim-overall-sim-response}
    \ddot{\myvec{\Phi}}(t) \triangleq\left(\prod_{m=M}^{2} {\ddot{\myvec{\Phi}}_m(t)} \myvec{\Xi}_m \right) {\ddot{\myvec{\Phi}}_1(t)}\in\mathbb{C}^{N_m \times N_m},
\end{equation}
where the input values may be encoded at the responses of $\ddot{\myvec{\Phi}}_1(t)$ and $\ddot{\myvec{\Phi}}_M(t)$ may play the role of the output vector at the \gls{RX}.

Since the latter operations are linear with respect to the impinging signal at each layer, and because every element in one layer contributes to the signal arriving at every element of the subsequent layer, this architecture bears a close resemblance to a \gls{DNN}'s fundamental building block: the fully connected linear layer.
Leveraging this resemblance, \glspl{D2NN} can be physically realized by treating the \gls{SIM} responses as the trainable weights of the network~\cite{XYN18, Qian2022_SIM_Rabbit}.

\begin{figure}[t]
    \centering
    \subfloat[A programmable MS acting as the input layer. The values of the input data vector (or matrix) are encoded as the phase responses of the MS elements and a directive antenna provides a beacon (i.e., non-information bearing) signal that illuminates the backplate of the first layer to initiate the forward pass.]{
        \begin{minipage}[t]{0.48\textwidth}
            \centering
            \includegraphics[width=\textwidth]{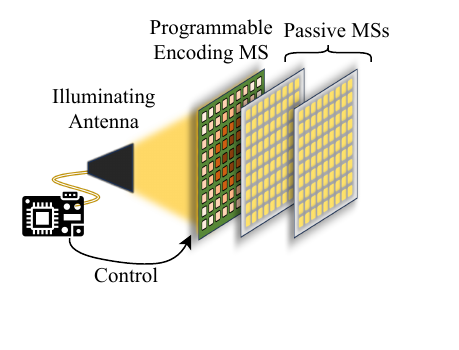}
        \end{minipage}
    }
    \hfill%
    \subfloat[An MS whose elements fully absorb the impinging signals acting as the output layer. The responses of the elements may be controllable and, therefore, adopted waveguide(s) at the backplate perform weighted summations on the received signals to reduce the resulting number of \gls{RF} chains needed for obtaining the output of the \gls{D2NN} in the digital domain.]{
        \begin{minipage}[t]{0.48\textwidth}
            \centering
            \includegraphics[width=\textwidth]{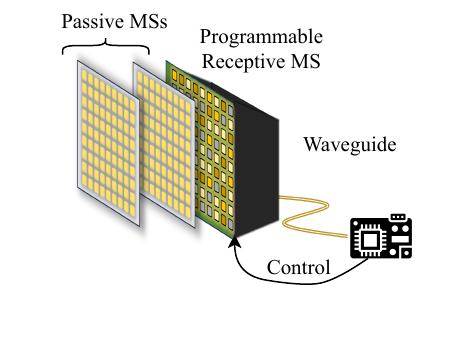}
        \end{minipage}
    }

    \caption{Illustration of techniques for transferring digital data to and from the RF domain using programmable \glspl{MS} as the first and last \gls{D2NN} layers. Once training is complete, unit cells characterized by the obtained responses for intermediate \gls{SIM} layers may be manufactured to be completely passive, providing important benefits in terms of energy consumption during wave computation.}
    \label{fig:sim_tx_rx}
\end{figure}

A prerequisite for \glspl{D2NN} is that the input data must be available in the \gls{RF} domain.
This requirement is inherently satisfied when the \gls{D2NN} is employed for processing ambient \gls{RF} signals in sensing applications. This offers significant advantages over digital \glspl{DNN} in terms of energy efficiency and latency, as it eliminates the need for analog to digital conversions.
Conversely, when a \gls{D2NN} performs inference on digital input data, the conversion to the \gls{RF} domain becomes crucial and requires complex hardware solutions at the first and last layers of the \gls{D2NN} as demonstrated in Fig.~\ref{fig:sim_tx_rx}.
The majority of the literature~\cite{LML23, Space_time_coding} achieves this conversion using programmable input layers within the \gls{SIM} structure, as depicted in Fig.~\ref{fig:sim_tx_rx}(a).
Each element of the digital input vector is mapped to the \gls{EM} response of a corresponding unit element in the initial layer, often through elementary techniques such as amplitude modulation.
Subsequently, a beacon signal, usually generated by a single directive antenna, illuminates the back of the first plate to initiate the forward \gls{DNN} pass.

The final stage, obtaining the \gls{D2NN} output, requires careful hardware design tailored to the target feature of the \gls{EI} process.
For classification problems, where the output maps to one or more predefined classes, signal receptors equal to the number of classes are placed after the final \gls{D2NN} layer. The predicted class index (in single class scenarios) corresponds to the receptor registering the highest observed signal strength.
For regression problems, obtaining the output is less straightforward. While it is possible to interpret the signal at the receptors as amplitude or phase modulated, achieving the requisite accuracy demands extremely fine grained beamforming capabilities from the \gls{SIM}, which may be impractical given current \gls{D2NN} hardware limitations.
An example is illustrated in Fig.~\ref{fig:sim_tx_rx}(b), where input receptors with potentially programmable responses are connected through waveguides to an \gls{RF} chain that combines the absorbed signals to produce the digital output.
Specifically during the training phase, once the forward pass is completed and the output is digitized via analog-to-digital converters contained in receive \gls{RF} chains, the result is compared against the expected target value for each training instance, and the loss function is computed digitally.
The backpropagation algorithm is then applied to calculate the necessary changes to the \gls{EM} responses of each \gls{SIM} unit element, and this iterative process continues until convergence.

Once the data are successfully converted to the \gls{RF} regime, \glspl{D2NN} execute computations at the speed of light, providing a decisive latency advantage over digital \glspl{DNN}.
Perhaps more critical are the benefits concerning power consumption.
Fabricated \glspl{MS} based on near passive circuitry, such as varactors, may operate on power levels as low as a few nanoWatts.
This represents a monumental improvement in energy efficiency compared to conventional \gls{DNN} processors, notably \glspl{GPU}, which often consume hundreds of Watts during inference.
Furthermore, once the optimal \gls{EM} responses for the \gls{MS} unit cells have been determined (typically via simulation), the \glspl{MS} can be manufactured to be entirely passive.
In such scenarios, the only power consuming components of \glspl{D2NN} are the feeding antenna and the output receptors, all of which can operate under very low power in controlled wireless environments where signal attenuation and multipath effects are minimal.
For applications where the input data are intrinsically in the \gls{RF} domain, such as those related to communications or sensing, the elimination of the power and latency overhead associated with analog-to-digital conversion establishes \glspl{D2NN} as an ideal candidate for innovative \gls{DNN} hardware solutions.

\subsection{Integration of D\textsuperscript{2}NNs in Wireless Systems}

Despite their inherent \gls{RF} nature, \glspl{D2NN} have historically been developed primarily as general purpose \gls{DNN} hardware accelerators, largely isolated from wireless system design.
To fully realize their promised advantages in \gls{EI} applications, the seamless integration of \gls{SIM}-based computing into the existing wireless communication infrastructure requires substantial research and development.
Crucially, the current design paradigm for \glspl{D2NN} is largely incompatible with existing and prospective wireless stacks.

A major concern revolves around the digital-to-analog conversion of input data.
Directly encoding each element of the digital input (e.g., an image pixel) as a pre-mapped \gls{EM} response of a programmable first \gls{SIM} layer faces severe limitations for practical deployment: (i) the power consumption of the \gls{MS} controller required for setting the responses can be considerable, (ii) achieving sufficient precision in programmable \gls{MS} responses to accurately encode high dimensional input data is technically challenging, and (iii) the physical size of the first \gls{MS} layer is directly proportional to the dimensionality of the input data.

Furthermore, this current practice fails to leverage the advanced capabilities of contemporary and future wireless systems.
Specifically, \gls{MIMO} systems can utilize transmission across multiple antennas to achieve spatial multiplexing and beamforming gains, thereby introducing new degrees of freedom for feeding input data into the \gls{SIM} device.
From this perspective, standard \gls{PHY} operations, including source encoding, modulation, and precoding, could be exploited to better integrate \glspl{D2NN} within \gls{MIMO} architectures.

Another vital consideration is the impact of the wireless channel on \gls{EI} performance.
Since \glspl{D2NN} were originally conceived for controlled wireless environments, such as free space, they rely on assumptions of high \gls{SNR} conditions and negligible, static wireless fading.
If \glspl{D2NN} are to be deployed for inference within wide area wireless networks, dynamic large- and small-scale fading effects can no longer be ignored.
In these realistic scenarios, the \gls{SIM} device assumes a dual functionality: Its learned \gls{EM} responses must not only execute the inference task, akin to \gls{DNN} layers, but also continuously adapt to the time varying channel conditions.
It is important to emphasize that the effects of the wireless channel should not solely be treated as a source of noise; instead, they can be harnessed as an additional computational resource, aligning with the \gls{OAC} paradigm~\cite{AirComp_Review}.

\section{Metasurfaces-Integrated Neural Networks}

Pondering on the aforementioned limitations of \glspl{D2NN} in their deployment on realistic wireless systems, it is evident that more advanced \gls{DNN} architectures are required that incorporate \gls{MS}-based wave computing practices within an \gls{E2E} system across the \gls{TX} and \gls{RX} devices, as well as the propagation environment. 
Such designs, dubbed \glspl{MINN} in the following~\cite{Stylianopoulos_GO}, are designed to also take advantage of wireless \gls{MIMO} systems and incorporate analog or digital \gls{DNN} processing at the transceivers.
The incorporated \gls{D2NN}(s) are split at the end devices or reside inside the wireless environment, offering \gls{OTA} computing with the inclusion of wireless fading~\cite{GJZ24_SIM_TOC, Stylianopoulos_GO}.
The intention of the \gls{MINN} system is to offload computation from the transceivers to the environment, allowing for benefits in terms of communication power and computational hardware complexity as well as processing energy.
The \gls{E2E} models presented in this section are designed to accommodate multiple variations in terms of physical devices (including \glspl{RIS} or \gls{SIM}), placement and capabilities thereof, as well as system objectives.

\subsection{System Model}
We begin by considering a more comprehensive \gls{MIMO} system compared to the straightforward one presented in~\eqref{eq:rayleigh-sommerfeld} of the previous section in order to account for arbitrary wireless fading and multi-antenna transceiver devices. We further assume the \gls{MS} device to be located within the environment.
Finally, to the intention of providing a unified system and a generic \gls{DNN} architecture, the system is designed to incorporate either a \gls{SIM} or \gls{RIS} device, so the term \gls{MS} is used, so as to be agnostic of the type of device, and a common notation is adopted for both cases, differentiating between the \gls{MS} types explicitly wherever needed.

\subsubsection{System and Received Signal Models}

We analyze the uplink of a point-to-point \gls{MIMO} communication system, where a \gls{TX} with $N_t$ antennas transmits data to an $N_r$-antenna \gls{RX} over time-indexed frames ($t=1,2,\ldots$). This communication is enhanced by an \gls{MS} (either an \gls{RIS} or \gls{SIM}), which operates as a standalone node. The \gls{MS} configuration can be changed at each discrete time step $t$ via an abstract controller~\cite{BAL24_RIS_review}. If the \gls{MS} is a \gls{SIM}, assume it comprises $M$ thin diffractive layers, each with $N_m$ elements, resulting in a total of $N \triangleq M N_m$ phase-tunable elements. For notational simplicity and to generalize the system model, we also use $N$ to denote the number of tunable elements for an \gls{RIS}.

Let $\mathbf{H}_{\rm D}(t) \in \mathbb{C}^{N_r \times N_t}$, $\mathbf{H}_{\rm 1}(t) \in \mathbb{C}^{N_t \times N_m}$, and $\mathbf{H}_{\rm 2}(t) \in \mathbb{C}^{N_r \times N_m}$ be the channel response matrices at time $t$ for the \gls{TX}-\gls{RX}, \gls{TX}-\gls{MS}, and \gls{MS}-\gls{RX} links, respectively. The transmitted signal is $\myvec{s}(t) \in \mathbb{C}^{N_t \times 1}$, constrained by a power budget $P\triangleq \mathbb{E}[\| \myvec{s}(t) \|]$. This signal vector encapsulates both the intended, source-coded, and modulated data stream (the number of data symbols $d\leq\min\{N_t,N_r\}$) and potential beamforming weights, without specific assumptions about its generation or symbol distribution.

\begin{figure}[t]
\centering
\includegraphics[width=0.6\linewidth]{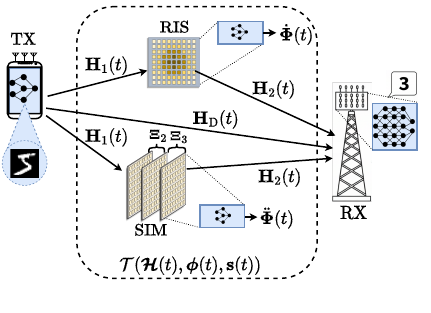}
\caption{The \gls{MIMO} system considered in the \gls{MINN} framework incorporating either an \gls{RIS} or a \gls{SIM} device. The \glspl{MS} may include a \gls{DNN}-based controller or a basic processing unit to store or update their fixed configuration.}
\label{fig:system_model}
\end{figure}

During each $t$-th frame transmission, the \gls{MS} is defined by its controllable \emph{phase configuration} vector $\boldsymbol{\omega}(t) \in\mathbb{C}^{N \times 1}$. The resulting \emph{response configuration}, assuming idealized unit amplitude, is $\boldsymbol{\phi}(t) \triangleq \exp(-\jmath \boldsymbol{\omega}(t))$. The collective effects of the metamaterial responses on the cascaded channel are captured by the matrix $\myvec{\Phi}(t) \in \mathbb{C}^{N_m \times N_m}$, the detailed structure of which is presented in the next subsection. Throughout this chapter, $\myvec{\Phi}(t)$, $\boldsymbol{\phi}(t)$, and $\boldsymbol{\omega}(t)$ serve as generic notation for both \gls{RIS} and \gls{SIM} cases, with device-specific notation introduced only where necessary.

The baseband received signal at the \gls{RX} antennas is thus given by:
\begin{align}
    \mathbf{y}(t)
    &\triangleq\left(\mathbf{H}_{\rm D}(t)+\mathbf{H}_{\rm 2}(t){\myvec{\Phi}}(t)\mathbf{H}_{\rm 1}^{\dagger} (t)\right)\myvec{s}(t) + \myvec{\tilde{n}}\label{eq:received-signal-RIS_1}\\
    &\triangleq \mathcal{T}\left(\boldsymbol{\mathcal{H}}(t), \boldsymbol{\phi}(t), \myvec{s}(t)\right)\label{eq:received-signal-RIS_2},
\end{align}
where $\myvec{\tilde{n}} \in \mathbb{C}^{N_r \times 1}$ is the \gls{AWGN} at the \gls{RX}, with \gls{iid} samples drawn from $\mathcal{CN}(0,\sigma^2)$. We utilize the transmission function $\mathcal{T}(\boldsymbol{\mathcal{H}}(t), \boldsymbol{\phi}(t), \myvec{s}(t))$ as an abstraction, highlighting that the wireless medium is treated as a programmable computation layer. In this formulation, $\boldsymbol{\mathcal{H}}(t) \triangleq \{\mathbf{H}_{\rm D}(t), \mathbf{H}_{\rm 1}(t), \mathbf{H}_{\rm 2}(t)\}$ denotes the instantaneous \gls{CSI}. This \gls{CSI} is assumed to be available to all system nodes. While this availability necessitates a challenging recurring channel estimation phase at each $t$-th step (see~\cite{Alexandropoulos2022Pervasive} and references therein), this assumption allows us to focus on the training and evaluation of the proposed \gls{MINN} architecture. Future work could integrate channel estimation into the \gls{DNN} transceiver modules following \gls{ISAC} principles~\cite{RIS_ISAC_SPM,RIS_ISAC_Chapter}. Alternatively, channel-agnostic transceiver variations are proposed and evaluated in subsequent sections to assess performance trade-offs when integrating \glspl{MS} as \gls{OTA} wave-domain neural network layers. The overall system model is shown in Fig.~\ref{fig:system_model}, illustrating the $\mathbf{H}_1(t)$ and $\mathbf{H}_2(t)$ links for both \gls{RIS}- and \gls{SIM}-enabled scenarios.

\subsubsection{RIS and SIM Models}\label{sec:ris-sim-models}
Starting with the \gls{RIS}, its phase configuration vector at time $t$ is denoted by $\dot{\boldsymbol{\omega}}(t) \triangleq [\dot{\omega}_1(t),\ldots,\dot{\omega}_{N_m}(t)]^\top$, which is equivalent to the generic notation $\boldsymbol{\omega}(t)$ introduced previously. The phase state of its $n$-th unit element ($n=1,2,\ldots,N_m$) is $\dot{\omega}_n(t) \in [0, 2\pi]$.
The induced response configuration vector is $\dot{\boldsymbol{\phi}}(t) \triangleq \exp(-\jmath \dot{\boldsymbol{\omega}}(t))$, equivalent to $\boldsymbol{\phi}(t)$ in~\eqref{eq:received-signal-RIS_2}. In this case, the response matrix is the diagonal matrix $\dot{\myvec{\Phi}}(t) \triangleq {\rm diag}(\dot{\boldsymbol{\phi}}(t))\in \mathbb{C}^{N_m \times N_m}$, so that $\dot{\myvec{\Phi}}(t)$ is equivalent to $\myvec{\Phi}(t)$ in \eqref{eq:received-signal-RIS_1}.

For the \gls{SIM} system model, we assume its $M$ layers ($m=1,2,\ldots,M$) are closely stacked and parallel, with their shared normal vector perpendicular to the \gls{TX}-\gls{RX} link. Under this arrangement, the \gls{TX} signal first hits the initial layer, undergoes diffraction and controllable phase shifting across the consecutive $M-1$ layers, and is then finally diffracted toward the \gls{RX}. 
Due to the compact spacing, the layer-to-layer propagation is accurately modeled by the Rayleigh-Sommerfeld diffraction equation provided in~\eqref{eq:rayleigh-sommerfeld}.

In addition to diffraction, each $n$-th element of the $m$-th \gls{SIM} layer introduces a controllable weight, $[\ddot{\boldsymbol{\phi}}_m(t)]_n \triangleq \exp(-\jmath \ddot{\omega}^m_n(t))$, where $\ddot{\omega}^m_n(t) \in [0, 2\pi]$ is the element's phase state. We define $\ddot{\boldsymbol{\phi}}_m(t)$ as the layer $m$'s response configuration, and $\ddot{\boldsymbol{\phi}}(t) \triangleq [\ddot{\boldsymbol{\phi}}^\top_1(t),\ldots,\ddot{\boldsymbol{\phi}}^\top_M(t)]^\top \in \mathbb{C}^{N \times 1}$, which is equivalent to $\boldsymbol{\phi}(t)$, as the overall response configuration.
The \gls{SIM} phase configuration vector is $\ddot{\boldsymbol{\omega}}(t) \triangleq [\ddot{\omega}^1_1(t), \ldots, \ddot{\omega}^M_{N_m}(t)]^\top \in \mathbb{C}^{N \times 1}$. 
The overall \gls{SIM} response matrix is mathematically expressed through~\eqref{eq:sim-overall-sim-response} as in the case of \glspl{D2NN}.

Note that $\myvec{\Phi}(t)\equiv \ddot{\myvec{\Phi}}(t)$ allows \eqref{eq:received-signal-RIS_1} to hold for the \gls{SIM} case.
Revisiting the generic notations, $\myvec{\Phi}(t)$ and $\boldsymbol{\omega}(t)$ are now specifically defined as $\myvec{\Phi}(t) \in \{ \dot{\myvec{\Phi}}(t), \ddot{\myvec{\Phi}}(t) \}$ and $\boldsymbol{\omega}(t) \in \{ \dot{\boldsymbol{\omega}}(t), \ddot{\boldsymbol{\omega}}(t) \}$ for the \gls{RIS} and \gls{SIM} respectively. In the rest of the chapter, $\myvec{\Phi}(t)$, $\boldsymbol{\omega}(t)$, and $\boldsymbol{\phi}(t)$ are used when the underlying operations are \gls{MS}-agnostic, whereas $\dot{\myvec{\Phi}}(t)$, $\ddot{\myvec{\Phi}}(t)$, and their associated vectors are utilized when differentiation between \glspl{RIS} and \gls{SIM} is required. The notation used to refer to the (generic) \gls{MS} and each of the \gls{RIS} and \gls{SIM} cases is summarized in Table~\ref{tab:glossary}.

\begin{table}[!t]
\caption{Symbols used to indicate generic \gls{MS} aspects and the explicit \gls{RIS} and \gls{SIM} variations.}
\centering
\label{tab:glossary}
\begin{tabular}{p{0.65\columnwidth} c}
\hline
\textbf{Description} & \textbf{Symbol} \\
\hline
MS phase configuration vector (generic) & $\boldsymbol{\omega}(t)$ \\
MS response configuration (generic) & $\boldsymbol{\phi}(t)$ \\
Effects of the MS response in the cascaded channel (generic)& $\myvec{\Phi}(t)$ \\
RIS phase state of the $n$-th element & $\dot{\omega}_n$ \\
RIS phase configuration vector & $\dot{\boldsymbol{\omega}}(t)$ \\
RIS response configuration & $\dot{\boldsymbol{\phi}}(t)$ \\
RIS effects in the cascaded channel & $\dot{\myvec{\Phi}}(t)$ \\
SIM phase state of the $n$-th unit element of the $m$-th layer & $\ddot{\omega}^m_n(t)$ \\
SIM phase configuration vector & $\ddot{\boldsymbol{\omega}}(t)$ \\
SIM response configuration vector for the $m$-th layer & $\ddot{\boldsymbol{\phi}}_m(t)$ \\
SIM response configuration & $\ddot{\boldsymbol{\phi}}(t)$ \\
SIM effects in the cascaded channel & $\ddot{\myvec{\Phi}}(t)$ \\
Output of the penultimate layer of the MS \textit{Controller} & $\hat{\boldsymbol{\omega}}$ \\
Trainable MS configuration (generic) & $\bar{\boldsymbol{\omega}}$ \\
Trainable MS response (generic)& $\bar{\boldsymbol{\phi}}$ \\
Trainable RIS configuration & $\bar{\boldsymbol{\omega}}_{\rm RIS}$ \\
Trainable RIS response & $\bar{\boldsymbol{\phi}}_{\rm RIS}$ \\
Trainable SIM configuration & $\bar{\boldsymbol{\omega}}_{\rm SIM}$ \\
Trainable SIM response & $\bar{\boldsymbol{\phi}}_{\rm SIM}$ \\
Trainable SIM's $m$-th layer effects in the cascaded channel & $\myvec{\bar{\Phi}}^m_{\rm SIM}$ \\
Trainable SIM's $m$-th layer response & $\bar{\boldsymbol{\phi}}^m_{\rm SIM}$ \\
\hline
\end{tabular}
\end{table}

\subsection{\gls{DNN} Architecture}

\begin{figure}[t]
\centering
    \includegraphics[width=\textwidth]{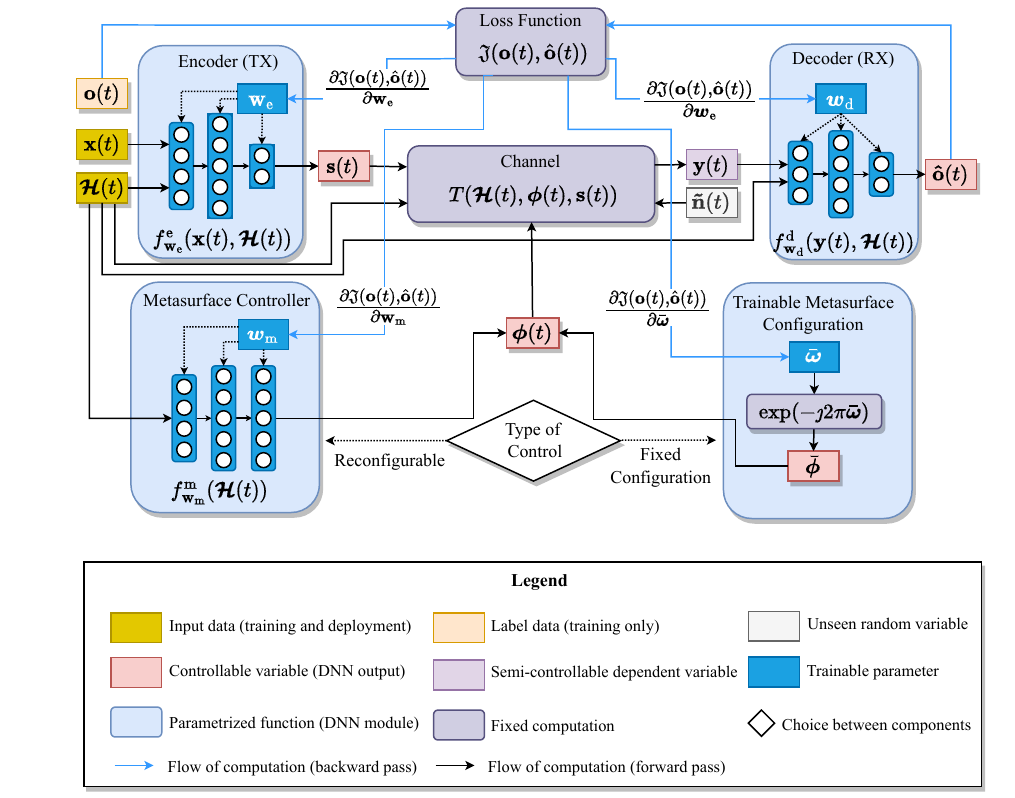}
    \caption{Block diagram and computation flow of the \gls{E2E} \gls{MINN} architecture~\cite{Stylianopoulos_GO} where the metasurface-parametrizable channel acts as an intermediate \gls{DNN} component. Both the cases of reconfigurable and static metasurfaces are included, entailing different procedures during the forward and backward passes.}
    \label{fig:architecture}
\end{figure}

The general architecture of the \gls{MINN} framework is consisted of three core modules, as depicted in Fig.~\ref{fig:architecture}.
The \textit{Encoder} and \textit{Decoder} modules are collocated with the transceivers and act similar to \gls{JSCC} and \gls{GOC} modules under the ``\gls{DNN}-splitting'' paradigm described earlier. They may be implemented either as fully digital \glspl{DNN}, running on edge \glspl{GPU}, or through analog hardware depending on the complexity of the layers and the capabilities of the devices.
The channel is treated as an intermediate module that incorporates uncontrollable computations induced by the channel fading and \gls{AWGN} in combination with controllable computations offered by the \gls{MS}.
The responses of the \gls{MS} may be treated similar to trainable \gls{DNN} parameters leading to static configurations, or they may be dynamically controlled through a \gls{MS} \textit{Controller} module that leads to reconfigurable instances.
The functionality of each module and the training procedure are detailed in the following sections.

\subsubsection{Transceiver Modules}
As introduced in Section~\ref{sec:EI-theory}, \gls{EI} requires two computational modules, implemented via digital processing hardware, collocated at the transceiver endpoints. To execute the \textit{infer-while-transmitting} methodology, the \gls{TX} employs an \textit{Encoder} \gls{DNN}, $f^{\rm e}_{\myvec{w}_{\rm e}}(\cdot)$, which produces the output $\myvec{s}(t)$, and the \gls{RX} utilizes a \textit{Decoder} \gls{DNN}, $f^{\rm d}_{\myvec{w}_{\rm d}}(\cdot)$, which generates the estimate $\hat{\myvec{o}}(t)$.
These modules are responsible for compression, encoding, decoding, error resilience and correction, as well as potential transmit and receive beamforming alongside probabilistic inference. The precise layer architecture of these models is intentionally left open, as the choice depends on several factors: (i) the characteristics of the wireless environment; (ii) the nature of the input and target data; (iii) the transceivers' hardware capabilities; and (iv) the current state-of-the-art. We note that different sub-modules may manage these operations, and typically, for uplink scenarios, $f^{\rm d}_{\myvec{w}_{\rm d}}(\cdot)$ can be implemented with larger \gls{DNN} structures due to the constant power supply available at base stations. Irrespective of the specific neural network chosen, a fixed post-processing step on the \textit{Encoder}'s output $\myvec{s}(t)$ is imposed to enforce the \gls{TX}'s power budget as follows:
\begin{equation}\label{eq:power-norm}
    \myvec{s}(t) \gets \sqrt{P}\frac{\myvec{s}(t)}{\|\myvec{s}(t)\|}.
\end{equation}

The concrete input arguments for the \textit{Encoder} and \textit{Decoder} functions lead to two variations, distinguished by the availability of \gls{CSI} at the endpoints.

\textbf{Channel-Agnostic Transceivers:}
In this variation, an instance of the data variable $\myvec{x}(t)$ is observed by the \gls{TX} and passed to the \textit{Encoder} to construct the transmitted signal. Concurrently, the \textit{Decoder} \gls{DNN} observes the received signal and estimates the unseen target variable $\myvec{o}(t)$:
\begin{align}
    \myvec{s}(t) &= f^{\rm e}_{\myvec{w}_{\rm e}}(\myvec{x}(t))\label{eq:encoder-channel-agnostc}, \\
    \hat{\myvec{o}}(t) &= f^{\rm d}_{\myvec{w}_{\rm d}}(\mathbf{y}(t)) \label{eq:decoder-channel-agnostc}.
\end{align}
Since \gls{CSI} is not utilized by the endpoints, this design resembles source-only coding, even though the \textit{Encoder} may incorporate redundancy, traditionally viewed as channel coding. Both processes must guarantee sufficient inference performance regardless of the current channel conditions, which is a demanding requirement. Nevertheless, this approach significantly simplifies the system architecture, and we include it in our subsequent investigations.

\textbf{Channel-Aware Transceivers:}
Assuming a quasi-static fading channel and a channel estimation procedure preceding data transmission in each $t$-th frame, the \gls{TX} and \gls{RX} modules obtain accurate estimates of the channel response matrices $\boldsymbol{\mathcal{H}}(t)$. Each module may use $\boldsymbol{\mathcal{H}}(t)$ as an additional input, yielding the following \textit{Encoder}/\textit{Decoder} \gls{DNN} representations:
\begin{align}
    \myvec{s}(t) &= f^{\rm e}_{\myvec{w}_{\rm e}}(\myvec{x}(t), \boldsymbol{\mathcal{H}}(t))\label{eq:encoder-channel-aware}, \\
    \hat{\myvec{o}}(t) &= f^{\rm d}_{\myvec{w}_{\rm d}}(\mathbf{y}(t), \boldsymbol{\mathcal{H}}(t)) \label{eq:decoder-channel-aware}.
\end{align}
Equipping the \gls{TX}/\gls{RX} modules with \gls{CSI} enables more resilient transmission schemes that closely resemble \gls{JSCC} \cite{GWT24_JSCC_Review}. The primary distinction is that \gls{JSCC} focuses on reconstructing $\myvec{x}(t)$, whereas \gls{EI} aims to approximate $\myvec{o}(t) = l(\myvec{x}(t))$.
Notice these two problems become equivalent by setting the mapping function $l(\cdot)$ to be the identity function, $\myvec{o}(t) = \myvec{x}(t)$, and adopting \gls{MSE} as the objective function $\mathfrak{J}(\cdot,\cdot)$. Therefore, \gls{EI} represents a more general problem formulation than communications, which typically concentrate on data reconstruction.
For the following, it is assumed that channel estimation is transparent and results in noise-free estimates of $\boldsymbol{\mathcal{H}}(t)$ before every transmission during both training and inference. Accounting for noisy estimates or integrating the estimation process within \gls{ISAC} paradigms constitutes open research directions.

\subsubsection{Control Module for Reconfigurable Metasurfaces}\label{sec:module-ms-controller}
When \gls{CSI} is available, which is common in many wireless communication contexts, the \gls{MS} dynamically adjusts its response configuration during each transmission frame to optimize the system's objective~\cite{10600711}. To integrate this dynamic capability into our \gls{E2E} architecture, the response $\boldsymbol{\phi}(t)$ is modeled as a controllable output of a third digital \gls{DNN}.
Specifically, the \gls{MS} \textit{Controller} is defined by the following neural network:
\begin{equation} \label{eq:ris-controller}
    \boldsymbol{\phi}(t) = f^{\rm m}_{\myvec{w}_{\rm m}}(\boldsymbol{\mathcal{H}}(t)),
\end{equation}
where the final layer operation enforces $\boldsymbol{\phi}(t) = \exp(-\jmath \hat{\boldsymbol{\omega}})$ with $\hat{\boldsymbol{\omega}}$ representing the output of the penultimate layer, with its elements constrained to the range $[0, 2 \pi]$. We maintain the abstract notation $\boldsymbol{\phi}(t)$, acknowledging that the specific output is either $\dot{\boldsymbol{\phi}}(t)$ or $\ddot{\boldsymbol{\phi}}(t)$ depending on the chosen \gls{MS} type. This perspective treats the \gls{MS} as an actively controlled entity capable of adapting to provide favorable wave-domain computation for every channel realization. This fine-grained control over environmental reprogrammability comes at the expense of an additional neural network module and its associated hardware requirements.

By substituting the outputs of the three trained modules from \eqref{eq:encoder-channel-agnostc}, \eqref{eq:decoder-channel-agnostc}, and \eqref{eq:ris-controller} into the received signal equation \eqref{eq:received-signal-RIS_1}, we obtain the full \gls{E2E} inference model $\hat{\myvec{o}}(t)=f^{\rm r}_{\myvec{w}_{\rm r}}(\myvec{x}(t), \boldsymbol{\mathcal{H}}(t))$ for both channel knowledge scenarios, as illustrated in Fig.\ref{fig:architecture}:

\begin{subequations}
\label{eq:e2e-model-reconfigurable}
\begin{align}
\hat{\myvec{o}}(t)&\!=\! \underbrace{f^{\rm d}_{\myvec{w}_{\rm d}} \bigg(\mathcal{T}\big(\boldsymbol{\mathcal{H}}(t), f^{\rm m}_{\myvec{w}_{\rm m}}(\boldsymbol{\mathcal{H}}(t)), f^{\rm e}_{\myvec{w}_{\rm e}}(\myvec{x}(t))\big)\bigg),}_{\text{channel-agnostic transceivers}} \label{eq:ris-control-module-csi-agnostic} \\
\hat{\myvec{o}}(t)&\!=\!
    \underbrace{f^{\rm d}_{\myvec{w}_{\rm d}} \bigg(\mathcal{T}\big(\boldsymbol{\mathcal{H}}(t), f^{\rm m}_{\myvec{w}_{\rm m}}(\boldsymbol{\mathcal{H}}(t)), f^{\rm e}_{\myvec{w}_{\rm e}}(\myvec{x}(t),\boldsymbol{\mathcal{H}}(t))\big), \boldsymbol{\mathcal{H}}(t)\bigg).}_{\text{channel-aware transceivers}} \label{eq:ris-control-module-csi-aware}
\end{align}
\end{subequations}
The entire set of trainable weights for this reconfigurable architecture is defined as $\myvec{w}_{\rm r} \triangleq \{\myvec{w}_{\rm d}, \myvec{w}_{\rm e}, \myvec{w}_{\rm m}\}$. These weights are jointly optimized using a shared objective function and backpropagation process, detailed in Section~\ref{subsec:Training}. Note that \eqref{eq:ris-control-module-csi-agnostic} deliberately allows the control module to be channel-aware. This implies that the \gls{MS} possesses sensing capabilities (e.g.,~\cite{10352433, ZWL23_Channel_Estimation}) to acquire channel knowledge.

\subsubsection{Metasurfaces with Trainable Static Response}\label{sec:ms-with-fixed-response}
An alternative, simplified approach is to directly learn a fixed response configuration for the \gls{MS}, which we denote by $\boldsymbol{\bar{\omega}}$. Although the training procedure involves optimizing $\boldsymbol{\bar{\omega}}$, the final learned configuration is fixed onto the \gls{MS} post-training. This results in a constant (static) response configuration $\boldsymbol{\phi}(t) \equiv \boldsymbol{\bar{\phi}} \triangleq \exp(-\jmath \boldsymbol{\bar{\omega}})$ that is maintained over time, regardless of channel conditions or input data. This paradigm treats the effective phase configurations similarly to \gls{DNN} weights, as they remain fixed after training and perform the same computational operation across varying input instances. The training procedure optimizes the combined weight set $\myvec{w}_{\rm s} \triangleq \{\myvec{w}_{\rm d}, \myvec{w}_{\rm e}, \bar{\boldsymbol{\omega}}\}$.

The \gls{E2E} static architecture is then expressed as follows:
\begin{subequations}
    \label{eq:e2e-model-static}
\begin{align}
\hat{\myvec{o}}(t) &= \underbrace{f^{\rm d}_{\myvec{w}_{\rm d}} \left(\mathcal{T}\left(\boldsymbol{\mathcal{H}}(t), \bar{\boldsymbol{{\phi}}}, f^{\rm e}_{\myvec{w}_{\rm e}}(\myvec{x}(t))\right)\right),}_{\text{channel-agnostic transceivers}} \\
    \hat{\myvec{o}}(t) &= \underbrace{f^{\rm d}_{\myvec{w}_{\rm d}} \bigg(\mathcal{T}\big(\boldsymbol{\mathcal{H}}(t), \bar{\boldsymbol{{\phi}}}, f^{\rm e}_{\myvec{w}_{\rm e}}(\myvec{x}(t),\boldsymbol{\mathcal{H}}(t))\big), \boldsymbol{\mathcal{H}}(t)\bigg).}_{\text{channel-aware transceivers}}
\end{align}
\end{subequations}
It is important to note that, while reconfigurable \glspl{MS} provide more granular control over shaping the transmission function $\mathcal{T}(\cdot)$, the inclusion of the \gls{MS} \textit{Controller} network may hinder the training efficiency of the proposed \gls{MINN} compared to the static variation. Furthermore, for wireless systems exhibiting limited variability, such as \gls{LoS}-dominant environments with fixed transceivers, a static \gls{MS} configuration may achieve satisfactory performance. The subsequent section addresses the systemic requirements for all variations presented here, with performance trade-offs explored in the numerical evaluations that follow.

\subsection{Training through Backpropagation on Wireless Channels}\label{subsec:Training}

To train the neural networks at the various wireless communication nodes, we first gather a labeled dataset $\mathcal{D} \triangleq \{ (\myvec{x}_i, \myvec{o}_i) \}_{i=1}^{|\mathcal{D}|}$ containing $|\mathcal{D}|$ data instances. Additionally, it is assumed access to a set of $|\mathcal{C}|$ channel sample estimates $\mathcal{C} \triangleq \{\boldsymbol{\mathcal{H}}(t) \}_{t=1}^{|\mathcal{C}|}$ observed at respective coherent time instances, which are not necessarily uniformly spaced. We make the critical assumption that the channel realizations are conditionally independent\footnote{In specific scenarios, the data realizations and channel statistics may be dependent. For instance, in target detection, where $\myvec{x}_i$ are sensory inputs and $\myvec{o}_i$ indicates target presence, deep fading may correlate with target blockages. In such cases, channel measurements and data observations must be collected synchronously, necessitating a more intricate \gls{EI} objective. However, the inference problem might be computationally simplified since \gls{CSI} provides supplementary information regarding the target value.} from the data instances in $\mathcal{D}$. This conditional independence permits the evaluation of the expectation in the $\mathcal{OP}_{\rm EI}$ objective function via \gls{iid} Monte Carlo sampling.

The training process is formulated as a variant of the standard gradient descent method for neural networks, extended to incorporate channel samples. We introduce the generic parameter vector $\myvec{w}_{\rm k}$, where ${\rm k} \in \{{\rm r},{\rm s}\}$ corresponds to the reconfigurable or static \gls{MS} selection, respectively. Similar to standard deep learning practices, our \gls{E2E} \gls{MINN} architecture can be optimized using \gls{SGD} over the collected data and channel instances.

The data-channel loss function is expressed as $\mathfrak{J}(\myvec{o}_i, \hat{\myvec{o}}_i) = \mathfrak{J}(\myvec{o}_i, f^{\rm k}_{\myvec{w}_{\rm k}}(\myvec{x}(t), \boldsymbol{\mathcal{H}}(t)))$, explicitly showing its dependence on the instantaneous wireless channel. Leveraging the conditional independence assumption, the $\mathcal{OP}_{\rm EI}$ objective can be approximated as follows:
\begin{equation}\label{eq:SGD-objective}
    \mathbb{E}_{\boldsymbol{\mathcal{H}}}[J(\myvec{w}_{\rm k})]\cong\frac{1}{|\mathcal{C}||\mathcal{D}|} \sum_{t=1}^{|\mathcal{C} |}    \sum_{i=1}^{|\mathcal{D}|} \mathfrak{J}(\myvec{o}_i, f^{\rm k}_{\myvec{w}_{\rm k}}(\myvec{x}_i, \boldsymbol{\mathcal{H}}(t))).
\end{equation}
In the online \gls{SGD} implementation, at each time $t$, a single data point and channel instance are selected to estimate the gradient, and the parameter vector is updated as:
\begin{equation}\label{eq:w-update}
    \myvec{w}_{\rm k} \gets \myvec{w}_{\rm k} - \eta \nabla_{\myvec{w}_{\rm k}}\mathfrak{J}(\myvec{o}(t), f^{\rm k}_{\myvec{w}_{\rm k}}(\myvec{x}(t), \boldsymbol{\mathcal{H}}(t))),
\end{equation}
for a chosen learning rate $\eta$. The gradient $\nabla_{\myvec{w}_{\rm k}}\mathfrak{J}$ is defined based on the \gls{MS} architecture:
\begin{align}
    \nabla_{\myvec{w}_{\rm k}}\mathfrak{J} &= \underbrace{\bigg[ \Big[\frac{\partial \mathfrak{J}}{\partial \myvec{w}_{\rm d}}\Big]^\top, \Big[\frac{\partial \mathfrak{J}}{\partial \myvec{w}_{\rm e}}\Big]^\top, \Big[\frac{\partial \mathfrak{J}}{\partial \myvec{w}_{\rm m}}\Big]^\top\bigg]^\top}_{\text{reconfigurable metasurface}}
    \\
    \nabla_{\myvec{w}_{\rm k}}\mathfrak{J} &= \underbrace{\bigg[ \Big[\frac{\partial \mathfrak{J}}{\partial \myvec{w}_{\rm d}}\Big]^\top, \Big[\frac{\partial \mathfrak{J}}{\partial \myvec{w}_{\rm e}}\Big]^\top, \Big[\frac{\partial \mathfrak{J}}{\partial \bar{\boldsymbol{{\omega}}}}\Big]^\top\bigg]^\top}_{\text{metasurface with trainable fixed response}}.
\end{align}
Under the \gls{iid} sampling assumption, consecutive gradient evaluations from \eqref{eq:w-update} provide unbiased estimators of the true gradient of \eqref{eq:SGD-objective}. Therefore, based on the stochastic approximation framework, repetitive application of this procedure converges to the true expected value with probability $1$ up to $O(\eta)$ precision when using a constant step size~\cite{Borkar2008}.

The complete training procedure supporting all variations (channel-agnostic/-aware transceivers, static/reconfigurable \gls{MS} controllers, and \gls{RIS}/\gls{SIM} structure) is detailed in Algorithm~\ref{alg:SGD}. Lines 6--9 implement the \gls{MINN} architecture as defined in \eqref{eq:e2e-model-reconfigurable} and \eqref{eq:e2e-model-static}. While more advanced optimization techniques such as batching, momentum, and adaptive rates \cite{Adam} can be used, they are omitted here for clarity.

\begin{algorithm}[t]
\caption{Training of the Proposed \gls{E2E} \gls{MINN}}
\label{alg:SGD}
\begin{algorithmic}[1]
\State Construct DNN weight vector $\myvec{w}_{\rm k}$ as one of the following:
\Statex \hspace{1em} \textit{i}) $\myvec{w}_{\rm k} = {\rm concat}(\myvec{w}_{\rm d}, \myvec{w}_{\rm e}, \myvec{w}_{\rm m})$. \hspace{1em} \Comment{$\myvec{w}_{\rm k} \gets \myvec{w}_{\rm r}$}
\Statex \hspace{1em} \textit{ii}) $\myvec{w}_{\rm k} = {\rm concat}(\myvec{w}_{\rm d}, \myvec{w}_{\rm e}, \bar{\boldsymbol{\omega}})$. \hspace{1em} \Comment{$\myvec{w}_{\rm k} \gets \myvec{w}_{\rm s}$}
\State Initialize $\myvec{w}_{\rm k}$ randomly.
\For{$t = 1, 2, \ldots, $ until convergence}
    \State Sample training data instance $(\myvec{x}(t), \myvec{o}(t))$ from $\mathcal{D}$.
    \State Sample random channel realization $\boldsymbol{\mathcal{H}}(t)$ from $\mathcal{C}$.
    \State Compute transmit signal $\myvec{s}(t)$ using one of the following:
    \Statex \hspace{2em} \textit{i}) $\myvec{s}(t) = f^{\rm e}_{\myvec{w}_{\rm e}}(\myvec{x}(t))$. \hspace{2em} \Comment{Eq.~\eqref{eq:encoder-channel-agnostc}}
    \Statex \hspace{2em} \textit{ii}) $\myvec{s}(t) = f^{\rm e}_{\myvec{w}_{\rm e}}(\myvec{x}(t), \boldsymbol{\mathcal{H}}(t))$. \hspace{2em} \Comment{Eq.~\eqref{eq:encoder-channel-aware}}
    \State Compute \gls{MS} response $\boldsymbol{\phi}(t)$ using one of the following:
    \Statex \hspace{2em} \textit{i}) $\boldsymbol{\phi}(t) = f^{\rm m}_{\myvec{w}_{\rm m}}(\boldsymbol{\mathcal{H}}(t))$. \hspace{2em} \Comment{Eq.~\eqref{eq:ris-controller}}
    \Statex \hspace{2em} \textit{ii}) $\boldsymbol{\phi}(t) = \bar{\boldsymbol{\phi}} = \exp(-\jmath \boldsymbol{\bar{\omega}})$.
    \State Transmit $\myvec{s}(t)$ to receive $\mathbf{y}(t)$:
    \Statex \hspace{2em} $\mathbf{y}(t) = \mathcal{T}(\boldsymbol{\mathcal{H}}(t), \boldsymbol{\phi}(t), \myvec{s}(t))$. \Comment{Eq.~\eqref{eq:received-signal-RIS_2}}
    \State Compute output $\hat{\myvec{o}}(t)$ using one of the following:
    \Statex \hspace{2em} \textit{i}) $\hat{\myvec{o}}(t) = f^{\rm d}_{\myvec{w}_{\rm d}}(\mathbf{y}(t))$. \hspace{2em} \Comment{Eq.~\eqref{eq:decoder-channel-agnostc}}
    \Statex \hspace{2em} \textit{ii}) $\hat{\myvec{o}}(t) = f^{\rm d}_{\myvec{w}_{\rm d}}(\mathbf{y}(t), \boldsymbol{\mathcal{H}}(t))$. \hspace{2em} \Comment{Eq.~\eqref{eq:decoder-channel-aware}}
    \State Update weights $\myvec{w}_{\rm k}$ through \gls{SGD} as:
    \Statex \hspace{2em} $\myvec{w}_{\rm k} \gets \myvec{w}_{\rm k}-\eta \nabla_{\myvec{w}_{\rm k}}\mathfrak{J}(\myvec{o}(t), f^{\rm k}_{\myvec{w}_{\rm k}}(\myvec{x}(t), \boldsymbol{\mathcal{H}}(t)))$. \Comment{Eq.~\eqref{eq:w-update}}
\EndFor
\State \Return $\myvec{w}_{\rm k}$
\end{algorithmic}
\end{algorithm}
The core of the training is the gradient update mechanism in \eqref{eq:w-update}. Since \eqref{eq:e2e-model-reconfigurable} and \eqref{eq:e2e-model-static} define differentiable operations with respect to $\myvec{w}_{\rm s}$ or $\myvec{w}_{\rm r}$, the partial derivatives are computed via automatic differentiation by applying the chain rule to the computational graph. For completeness, we provide the derivations for the partial derivatives of the various modules, treating the implementation-defined derivatives of the classical neural network components (i.e., $\partial f_{\myvec{w}_{\mathrm{e}}}^{\mathrm{e}}/\partial \myvec{w}_{\mathrm{e}}$, $\partial f_{\myvec{w}_{\mathrm{d}}}^{\mathrm{d}}/\partial \myvec{w}_{\mathrm{d}}$, $\partial f_{\myvec{w}_{\mathrm{m}}}^{\mathrm{m}}/\partial \myvec{w}_{\mathrm{m}}$, and $\partial f_{\myvec{w}_{\mathrm{d}}}^{\mathrm{d}} / \partial \mathbf{y}(t)$) as known.

\subsubsection{Reconfigurable Metasurface (\gls{RIS} or \gls{SIM})}
For the reconfigurable \gls{MS} case, $\hat{\myvec{o}}$ is given by \eqref{eq:e2e-model-reconfigurable}. Applying backpropagation yields the following derivatives:
\begin{align}
    \frac{\partial \mathfrak{J}}{\partial \myvec{w}_{\rm d}}
    &= \frac{\partial \mathfrak{J}}{\partial \hat{\myvec{o}}(t)} \frac{\partial f_{\myvec{w}_{\rm d}}^{\rm d}}{\partial \myvec{w}_{\rm d}}, \label{eq:backprop_d} \\
    \frac{\partial \mathfrak{J}}{\partial \myvec{w}_{\rm m}}
    &= \frac{\partial \mathfrak{J}}{\partial \hat{\myvec{o}}(t)} 
    \frac{\partial f_{\myvec{w}_{\rm d}}^{\rm d}}{\partial \mathbf{y}(t)}
    \frac{\partial  \mathbf{y}(t)}{\partial f^{\rm m}_{\myvec{w}_{\rm m}}}
    \frac{\partial  f^{\rm m}_{\myvec{w}_{\rm m}}}{\partial \myvec{w}_{\rm m}}, \label{eq:backprop_m} \\
    \frac{\partial \mathfrak{J}}{\partial \myvec{w}_{\rm e}}
    &= \frac{\partial \mathfrak{J}}{\partial \hat{\myvec{o}}(t)} 
    \frac{\partial f_{\myvec{w}_{\rm d}}^{\rm d}}{\partial \mathbf{y}(t)} 
    \frac{\partial \mathbf{y}(t)}{\partial f_{\myvec{w}_{\rm e}}^{\rm e}} 
    \frac{\partial f_{\myvec{w}_{\rm e}}^{\rm e}}{\partial \myvec{w}_{\rm e}} \label{eq:backprop_e},
\end{align}
where $\partial \mathfrak{J}/\partial \hat{\myvec{o}}(t)$ is the gradient of the loss function with respect to the network's output (e.g., $-\myvec{o}(t)/\hat{\myvec{o}}(t)$ for \gls{CE} loss). The remaining partial derivatives involving the channel are:
\begin{align}
    \frac{\partial \mathbf{y}(t)}{\partial f_{\myvec{w}_{\rm e}}^{\rm e}} &= \mathbf{H}_{\rm 2}(t)\myvec{\Phi}(t)\mathbf{H}_{\rm 1}^{\dagger} (t) + \mathbf{H}_{\rm D}(t), \\
    \frac{\partial \mathbf{y}(t)}{\partial f^{\rm m}_{\myvec{w}_{\rm m}}} &=\frac{\partial \mathbf{y}(t)}{\partial \boldsymbol{\phi}(t)} = \big((\myvec{s}^{\top}(t)\mathbf{H}_{\rm 1}^{\ast}(t)) \otimes \mathbf{H}_{\rm 2}(t)\big)\myvec{D} \label{eq:phi-gradient}.
\end{align}

\subsubsection{\gls{RIS} with Fixed Configuration}
For the fixed-configuration \gls{RIS} case, $\hat{\myvec{o}}$ is computed via \eqref{eq:e2e-model-static}. Let $\boldsymbol{\bar{\omega}}_{\rm RIS} \in [0, 2\pi)^{N_m}$ and $\boldsymbol{\bar{\phi}}_{\rm RIS} \triangleq \exp(-\jmath \boldsymbol{\bar{\omega}}_{\rm RIS})$. Since $\partial \mathfrak{J}/\partial \myvec{w}_{\rm d}$ and $\partial \mathfrak{J}/\partial \myvec{w}_{\rm e}$ remain the same as in \eqref{eq:backprop_d} and \eqref{eq:backprop_e}, we focus on the derivative with respect to the trainable configuration $\boldsymbol{\bar{\omega}}$:
\begin{align}
    \frac{\partial \mathfrak{J}}{\partial \boldsymbol{\bar{\omega}}}
    = \frac{\partial \mathfrak{J}}{\partial \boldsymbol{\bar{\omega}}_{\rm RIS}}
    &=  \frac{\partial \mathfrak{J}}{\partial \hat{\myvec{o}}(t)}
    \frac{\partial f_{\myvec{w}_{\rm d}}^{\mathrm{d}}}{\partial \mathbf{y}(t)}
    \frac{\partial \mathbf{y}(t)}{\partial \boldsymbol{\bar{\phi}}_{\rm RIS}}
    \frac{\partial \boldsymbol{\bar{\phi}}_{\rm RIS}}{\partial \boldsymbol{\bar{\omega}}_{\rm RIS}},
    \label{eq:backprop_ris}
\end{align}
where $\partial \mathbf{y}(t)/\partial \boldsymbol{\bar{\phi}}_{\rm RIS}$ is computed via \eqref{eq:phi-gradient}, and $\partial \boldsymbol{\bar{\phi}}_{\rm RIS} / \partial \boldsymbol{\bar{\omega}}_{\rm RIS} = -\jmath \exp{(-\jmath \boldsymbol{\bar{\omega}}_{\rm RIS})}$.

\subsubsection{Fixed-Configuration \gls{SIM}}
For the fixed configuration \gls{SIM} case, $\hat{\myvec{o}}$ is again computed via \eqref{eq:e2e-model-static}. We use the notations $\boldsymbol{\bar{\omega}}_{\rm SIM} \in [0, 2\pi)^{N}$ and $\boldsymbol{\bar{\phi}}_{\rm SIM} \triangleq \exp(-\jmath \boldsymbol{\bar{\omega}}_{\rm SIM})$. The gradient is:
\begin{align}
    \frac{\partial \mathfrak{J}}{\partial \boldsymbol{\bar{\omega}}}
    = \frac{\partial \mathfrak{J}}{\partial \boldsymbol{\bar{\omega}}_{\rm SIM}}
    &=  \frac{\partial \mathfrak{J}}{\partial \hat{\myvec{o}}(t)}
    \frac{\partial f_{\myvec{w}_{\rm d}}^{\mathrm{d}}}{\partial \mathbf{y}(t)}
    \frac{\partial \mathbf{y}(t)}{\partial  \boldsymbol{\bar{\phi}}_{\rm SIM}}
    \frac{\partial \boldsymbol{\bar{\phi}}_{\rm SIM}}{\partial \boldsymbol{\bar{\omega}}_{\rm SIM}},
    \label{eq:backprop_sim}
\end{align}
where, $\partial \boldsymbol{\bar{\phi}}_{\rm SIM}/\partial \boldsymbol{\bar{\omega}}_{\rm SIM} = -\jmath \exp{(-\jmath \boldsymbol{\bar{\omega}}_{\rm SIM})}$. Since $\mathcal{T}(\cdot)$ now incorporates the \gls{SIM} model from \eqref{eq:sim-overall-sim-response}, calculating $\partial \mathbf{y}(t) / \partial \boldsymbol{\bar{\phi}}_{\rm SIM}$ requires further derivation. By denoting the response matrix of the $m$-th \gls{SIM} layer as $\myvec{\bar{\Phi}}^m_{\rm SIM} \triangleq {\rm diag}( \boldsymbol{\bar{\phi}}^m_{\rm SIM})$, where $\boldsymbol{\bar{\phi}}^m_{\rm SIM}$ is the trainable response of the $m$-th layer, we find $\partial \mathbf{y}(t) / \partial  \boldsymbol{\bar{\phi}}_{\rm SIM} = [[\partial \mathbf{y}(t) / \partial  \boldsymbol{\bar{\phi}}^1_{\rm SIM}]^\top, \ldots, [\partial \mathbf{y}(t) / \partial  \boldsymbol{\bar{\phi}}^M_{\rm SIM}]^\top]^\top$ with:
\begin{equation}\label{eq:phi-gradient-sim}
\frac{\partial \mathbf{y}(t)}{\partial  \boldsymbol{\bar{\phi}}^m_{\rm SIM}} =
\begin{cases}
    \begin{array}{l}
        \left(\myvec{s}^{\top}(t)\mathbf{H}_{\rm 1}^{\ast}(t)\right)
        \otimes \left(\mathbf{H}_{\rm 2}(t) \prod\limits_{m'=M}^{2} \myvec{\bar{\Phi}}^{m'}_{\rm SIM} \myvec{\Xi}_{m'}\right)\myvec{D},
    \end{array} & m=1 \\[2ex]
    \begin{array}{l}
        \left(  \left(\prod\limits_{m'=m}^{2} \myvec{\Xi}_{m'} \myvec{\bar{\Phi}}^{m'-1}_{\rm SIM}  \right) \mathbf{H}_{\rm 1}^{\dagger}(t)\myvec{s}(t)\right)^\top  \\
        \otimes(\mathbf{H}_{\rm 2}(t) \prod\limits_{m'=M}^{m+1} \myvec{\bar{\Phi}}^{m'}_{\rm SIM} \myvec{\Xi}_{m'} )\myvec{D},
    \end{array} & 1 < m \leq M.
\end{cases}
\end{equation}

In the above, we utilize the identity $\textrm{vec}(\mathbf{A} \mathbf{X} \mathbf{B}) = (\mathbf{B}^\top \otimes \mathbf{A}) \textrm{vec}(\mathbf{X})$ and defined the diagonal selection matrix $\myvec{D} \in [0,1]^{N_m^2 \times N_m}$ such that, for a vector $\myvec{x}$ and $\myvec{X} = {\rm diag}(\myvec{x})$, ${\rm vec}(\myvec{X}) = \myvec{D}\myvec{x}$.

\section{Image Classification at the Edge}

\begin{figure}[t]
    \centering
    \includegraphics[width=0.9\linewidth]{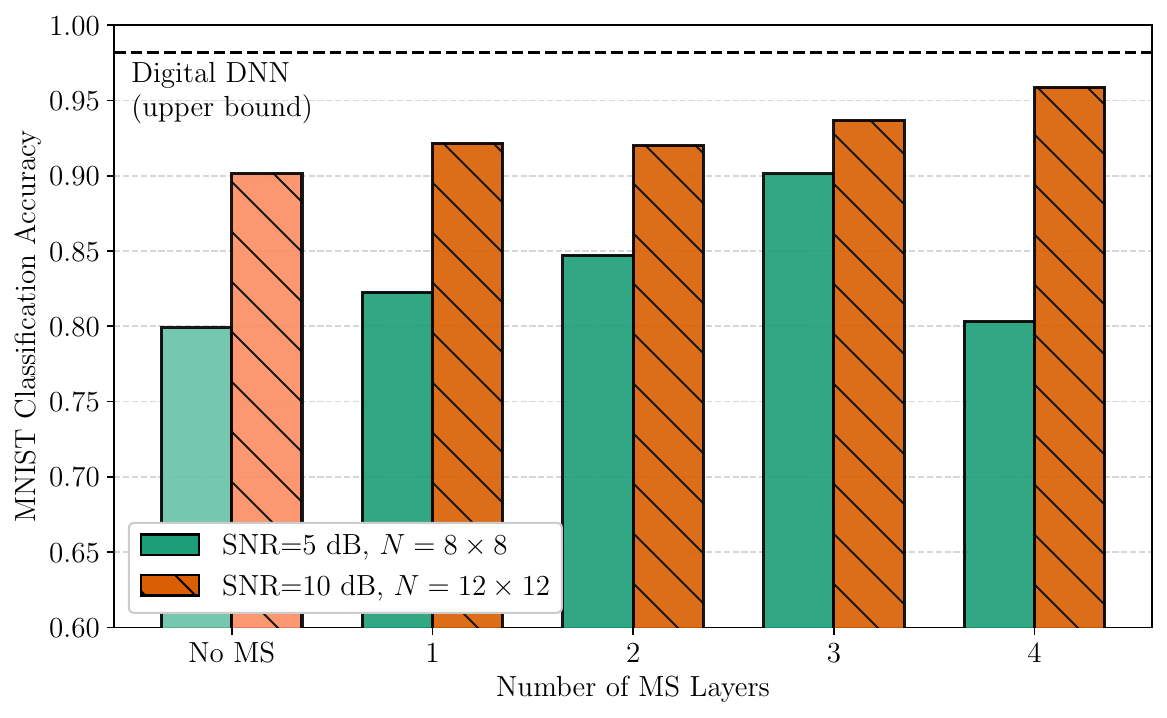}
    \caption{Mean classification accuracy of \gls{MINN} on static fading considering \glspl{RIS} and \gls{SIM} with different numbers of elements and layer, as well as two transmit \gls{SNR} values.}
    \label{fig:results-static-fading}
\end{figure}

In this section, the performance of the presented \gls{MINN} framework for the case of multi-class classification on the widely used MNIST data set~\cite{MNIST} is demonstrated.
This data set contains $60,000$ grayscale images of $28\times28$ pixels, each one depicting a handwritten digit from $0$ to $9$, and the $10$ classes correspond to the numerical value of each depicted digit.
We start by considering static fading, i.e., all three matrices $\mathbf{H}_{\rm D}$, $\mathbf{H}_{\rm 1}$, and $\mathbf{H}_{\rm 2}$
of $\boldsymbol{\mathcal{H}}$ are sampled using the Saleh-Valenzuela geometric model~\cite{SV_Model}, using $10$ scattering points that remain fixed throughout the training and inference scenario.
For our \gls{MINN} system based on a $4\times4$ \gls{MIMO} with noise variance $\sigma^2=-90$ dBm, we incorporate an \textit{Encoder} and a \textit{Decoder} module at the \gls{TX} and \gls{RX}, respectively, with three linear layers each and without channel knowledge.
The \gls{RIS} or the \gls{SIM} have their phase configurations directly trainable without the use of a \textit{Controller} module.
Training was conducted for $150$ epochs, using the Adam~\cite{Adam} variation of \gls{SGD} with a learning rate of $10^-3$ and its respective $\beta_1$ and $\beta_2$ parameters that control the momentum of the updates were set to $0.9$ and $0.999$, respectively.

As it can be observed from Fig.~\ref{fig:results-static-fading}, the combination of increased transmit \gls{SNR} and greater numbers of elements in the \glspl{MS} provides substantial benefits in terms of classification accuracy. 
It is noted that deeper \gls{SIM} structures are not always more efficient: During the propagation at the \gls{SIM} layers the signal attenuates with $O(d^3_{n,n'})$ as it can be observed from~\eqref{eq:rayleigh-sommerfeld}, which may result to weak signals at the \gls{RX}, negating the benefits of the additional computation brought by each consecutive \gls{SIM} in the low \gls{SNR} regime.
On the contrary, under sufficient \gls{SNR}, deeper \gls{SIM} architectures can be effective computational models and even approach the performance of the fully digital benchmark.
This \gls{DNN} contains digital layers of the same number and sizes as the \textit{Encoder} and \textit{Decoder} combined and performs inference directly, without accounting for any channel effects, i.e., $\mathcal{T}(\cdot)$ may be thought of as an identity layer.

\begin{figure}[t]
    \subfloat[$N_t=4$\label{fig:minn-clf-a}]{%
        \includegraphics[width=0.9\columnwidth]{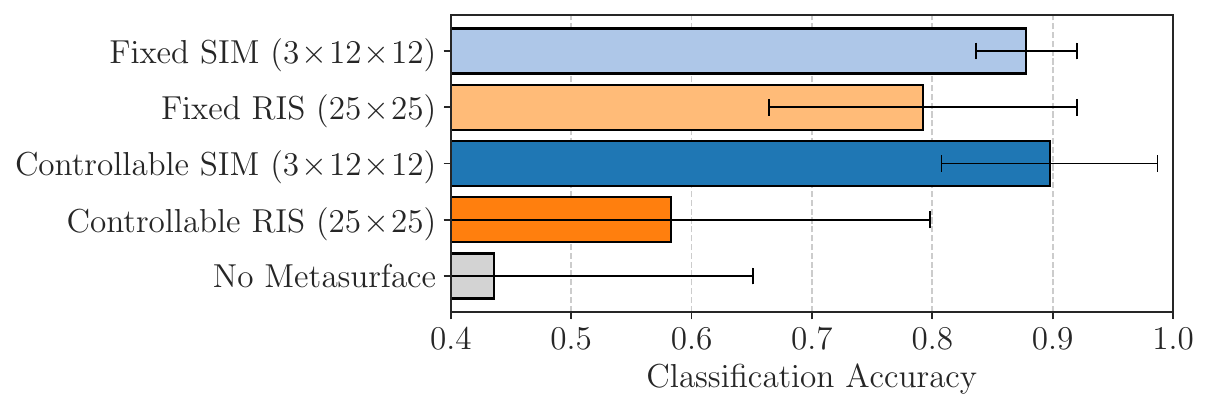}
    }

    \subfloat[$N_t=12$\label{fig:minn-clf-b}]{%
        \includegraphics[width=0.9\columnwidth]{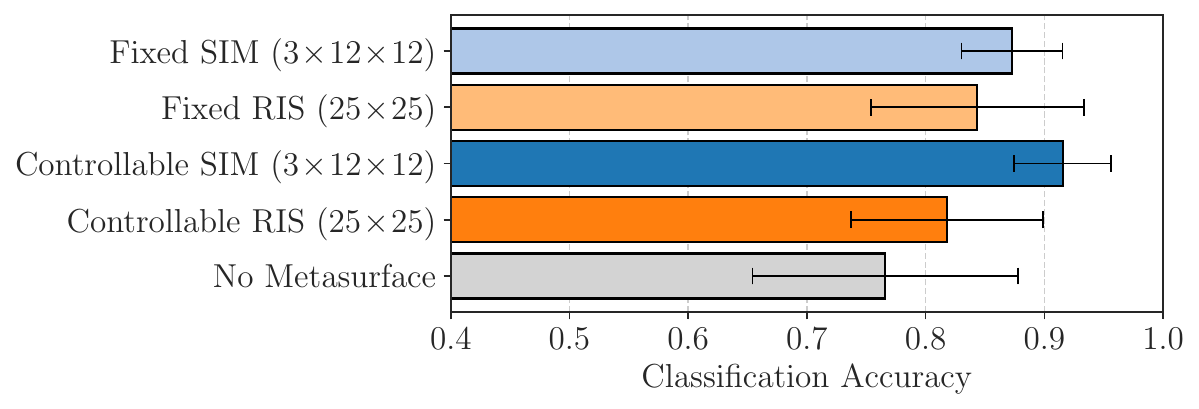}
    }

    \caption{Mean classification accuracy and standard deviations on the MNIST data set using dynamic Ricean fading under different number of \gls{TX} antennas and different \gls{MINN} architectures.}
    \label{fig:minn-contr-vs-static}
\end{figure}

Proceeding, we investigate the role of the \textit{Controller} module that offers dynamic reconfiguration on the employed \glspl{MS} instead of relying on static learned responses.
To do so, we now adopt dynamic Ricean fading following the model described in detail in~\cite{Alexandropoulos2022Pervasive}, so that the \gls{CSI} changed at every time step $t$.
The Ricean factors (in dB) for the \gls{TX}-\gls{MS}, \gls{MS}-\gls{RX}, and \gls{TX}-\gls{RX} links were respectively set to $13$, $7$, and $3$, providing a case of moderately to strongly dominant line of sight component.
To exploit this information, we employ channel-aware \textit{Encoder} and \textit{Decoder} modules as such: In each of these modules, a separate branch of layers receives as inputs the concatenated instantaneous channel matrices $\mathbf{H}_{\rm D}(t)$, $\mathbf{H}_{\rm 1}(t)$, and $\mathbf{H}_{\rm 2}(t)$ and extracts features to an arbitrary intermediate vector $\boldsymbol{v}(t)$.
The \textit{Encoder} and \textit{Decoder} modules parse their respective inputs ($\mathbf{x}(t)$ and $\mathbf{y}(t)$) through separate branches of three fully connected layers, the outputs of which are concatenated with $\boldsymbol{v}(t)$ and the resulting intermediate vector is passed through the final two fully connected layers to obtain the respective outputs ($\mathbf{s}(t)$ and $\hat{\mathbf{o}}(t)$). The \textit{Controller} module uses four additional fully connected layers that receive $\boldsymbol{v}(t)$ as input and output the desired \gls{MS} configuration $\boldsymbol{\phi}(t)$.

To sufficiently train the \gls{MINN} variations under dynamic fading on the MNIST data set, the next experiments use $1000$ epochs with a transmission power $P=30$ dBm and $N_r=32$. Results are averaged over $10$ different initializations.
The bars of Fig.~\ref{fig:minn-contr-vs-static} depict the mean accuracy under different numbers of \gls{TX} antennas with their corresponding standard deviations.
\gls{MINN} variations with a Control module are referred to as ``Controllable'', while the term ``Fixed'' corresponds to \glspl{MINN} that have their \gls{MS} configurations directly trainable, and therefore remain static during inference regardless of the \gls{CSI}.
It can first be inferred that the use of the Control module offers a slight advantage at the scenario at hand of around $3\%$ accuracy improvement for the \gls{SIM} cases.
The increase in the available link budget when $N_t=12$ substantially increases the performance of the comparatively less powerful \gls{RIS} devices and the baseline where \gls{EI} is performed solely through the \textit{Encoder} and the \textit{Decoder} \glspl{DNN} without the inclusion of any \gls{MS}. As expected, the variance across restarts is noticeably reduced.
In more dynamic scenarios that incorporate user and/or background scatter mobility, the improvements brought by the \textit{Controller} module are expected to be more pronounced.

\begin{figure}[t]
    \centering
    \includegraphics[width=0.9\linewidth]{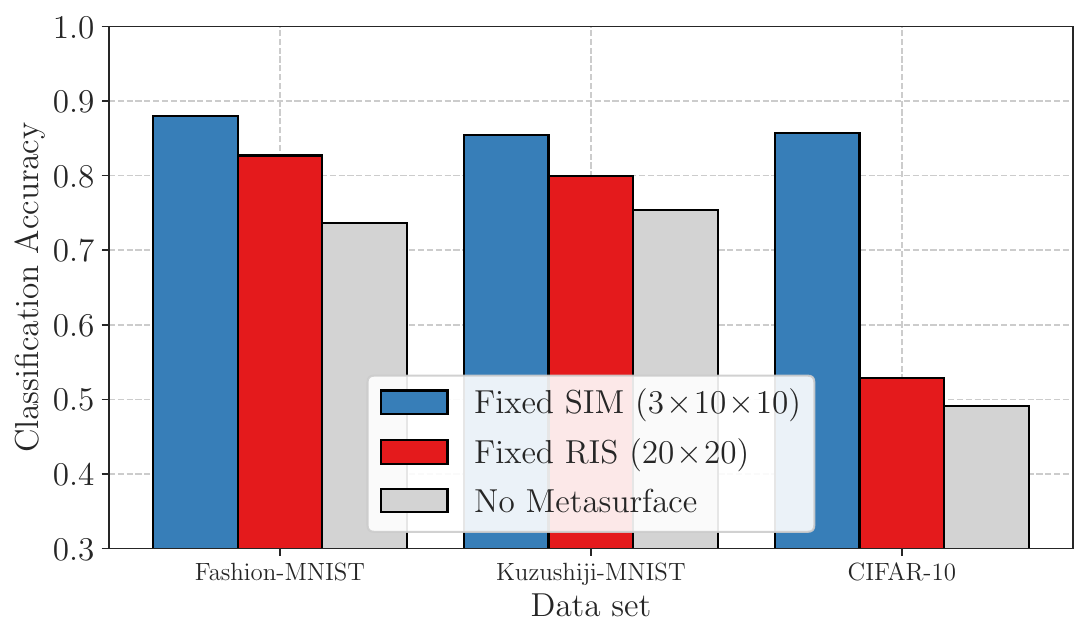}
    \caption{Mean accuracy with different \gls{MINN} variations and the No-\gls{MS} baseline, considering different data sets, fixed configuration \glspl{MS}, and channel-agnostic transceivers.}
    \label{fig:other-datasets}
\end{figure}

Next, the generalization of the \gls{MINN} framework is investigated over multiple data sets. The Fashion and Kuzushiji MNIST variations have been considered that depict grayscale images of clothing and historical Japanese characters, as well as the CIFAR-10 data set to provide increasing degrees of difficulty. For the two MNIST variations, we have replaced the first three layers of the \textit{Encoder} that parse $\mathbf{x}(t)$ with convolutional layers followed by max-pooling operations, while a deeper convolutional \textit{Encoder} \gls{DNN} architecture with skip connections has been used for the challenging case of CIFAR-10.
The performance across those data sets is given in Fig.~\ref{fig:other-datasets}, where it can be seen that the \gls{SIM}-based \gls{MINN} outperforms the \gls{RIS} variation and the baseline, especially under the case of CIFAR-10.

Since the intention of \glspl{MINN} is to offload computations onto the wireless channels in order to simplify transceiver hardware, we further elaborate on the energy consumption benefits of the presented methodologies.
We compare against the two other methodologies presented in Section~\ref{sec:EI-theory}.
In the ``infer-then-transmit'' paradigm, the classification is solved at the \gls{TX} using the ``Digital DNN'' benchmark discussed earlier for MNIST as well as the \textit{Encoder} architecture for CIFAR-10 with three more layers added for final classification.
A perfect transmission of the result is assumed, since it requires mere $\log_2(10) \approx 3.32$ bits.
For the ``transmit-then-infer'' case, we compress the input images using a compression scheme based on pre-trained \glspl{AE}, and we transmit the encoded representations using \gls{PSK} modulation and capacity achieving \gls{MIMO} precoding and combining. The ``Digital DNN'' is thus placed at the \gls{RX} and classifies the reconstructed images, achieving up to $80\%$ accuracy due to the limited transmission budget.
The power required for the computation of the \gls{TX} processing $P_{\rm net}$ is measured experimentally in the workstation used to run the experiments for each case, while the latency $\tau_{\rm inf}$ associated with performing inference over the test set is also recorded. The consumed energy per data instance is therefore calculated as $E = P_{\rm net} / \tau_{\rm inf}$. Those numbers are presented in Table~\ref{tab:energy_consumption}. It can be seen that the \gls{MINN} system architecture consumes energy comparable to the ``transmit-then-infer'' method which contains only the minimum processing for transmission, while it maintains an around $10\%$ performance improvement in terms of classification accuracy, motivating the \gls{OTA} methodology developed in this chapter.

\begin{table}[t]
    \centering
    \caption{Estimated computational energy consumption for the \textit{Encoder} device at the \gls{TX} under the three EI paradigms.}
    \label{tab:energy_consumption}
    \begin{tabular}{|l|c|c|c|}
        \hline
        \multicolumn{4}{|c|}{\textbf{MNIST}} \\
        \hline
        \textbf{Strategy} & $P_{\rm net} ({\rm W})$ & $\tau_{\rm inf} ({\rm s})$ & $E ({\rm mJ}/{\rm inst.})$ \\
        \hline
        Infer-then-transmit & $41.97$ & $8.55$ & $5.98$ \\
        Transmit-then-infer & $44.43$ & $3.27$ & $2.42$ \\
        Infer-while-transmitting (MINN) & $39.61$ & $3.62$ & $2.39$ \\
        \hline
        \multicolumn{4}{|c|}{\textbf{CIFAR-10}} \\
        \hline
        \textbf{Strategy} & $P_{\rm net} ({\rm W})$ & $\tau_{\rm inf} ({\rm s})$ & $E ({\rm mJ}/{\rm inst.})$ \\
        \hline
        Infer-then-transmit & $89.47$ & $10.63$ & $15.85$ \\
        Transmit-then-infer & $43.04$ & $3.68$ & $2.64$ \\
        Infer-while-transmitting (MINN) & $43.82$ & $5.30$ & $3.87$ \\
        \hline
    \end{tabular}
\end{table}

\section{Transmission Power Control}
Having showed that \glspl{MINN} may offer accuracy approaching that of fully digital \glspl{DNN} with reduced computational power requirements at the \gls{TX} side, it raises the question whether power savings for data transmission may also be achievable.
As illustrated in Fig.~\ref{fig:results-static-fading}, higher transmission power indeed offers better inference accuracy. However, the previous \gls{MINN} architectures were trained in fixed, high-\gls{SNR} conditions, which did not incentivize training to generalize to low-\gls{SNR} regimes.

To specifically treat this limitation, a \gls{MINN} variation is developed next, where the \gls{E2E} system learns to control the transmission power dynamically during training.
This is particularly designed for dynamic fading scenarios where one of the endpoints (typically, the \gls{RX} in downlink communications) is mobile, and therefore the required power level depends on the dynamic fading conditions of the instantaneous links.
In doing so, let us consider a modified version of the \gls{EI} problem~\cite{Stylianopoulos_MINN_Power}.
Based on a collected data set $\mathcal{D}$ and corresponding channel realizations $\boldsymbol{\mathcal{H}}$, the \gls{MINN} is seen as a mapping function $ \{ \mathbf{\hat{o}}{(t)}, P{(t)} \} \triangleq f_{{\bar{\mathbf{w}}}}(\mathbf{x}{(t)}, \boldsymbol{\mathcal{H}}{(t)})$, parametrized by a weight vector $\bar{\mathbf{w}}$, that outputs an estimate $\mathbf{\hat{o}}{(t)}$ for $\mathbf{o}{(t)}$ as well as the \gls{TX} power $P{(t)}$.
We thereby propose the following problem formulation for the design of $\bar{\mathbf{w}}$:
\begin{subequations}\label{eq:problem-power-control}
\begin{align}
\min_{\bar{\mathbf{w}}}\mathcal{L}(\bar{\mathbf{w}}) &\triangleq \mathbb{E}_{\boldsymbol{\mathcal{H}}}\left[ \frac{1}{|\mathcal{D}|}\sum_{i=1}^{|\mathcal{D}|}\mathfrak{J}_{\rm CE}\left(\mathbf{\hat{o}}{(i)}, \mathbf{o}{(i)}\right)\right], \label{eq:objective} \\
{\rm s.t.}~~& \{ \mathbf{\hat{o}}{(i)}, P{(i)} \} = f_{{\bar{\mathbf{w}}}}(\mathbf{x}{(i)}, \boldsymbol{\mathcal{H}}{(i)}), \\
&\mathbb{E}_{\boldsymbol{\mathcal{H}}}[P{(i)}] \leq P_{\rm max}~~\forall i\!=\!1,\dots,|\mathcal{D}|, \label{eq:power-constrint}
\end{align}
\end{subequations}
where $\mathbb{E}_{\boldsymbol{\mathcal{H}}}\left[\cdot\right]$ represents expectation over the channels, which is computed empirically as in~\eqref{eq:SGD-objective}.
Once problem~\eqref{eq:problem-power-control} is solved, $f_{\bar{\mathbf{w}}}(\cdot)$ may be used to infer the target values for previously unseen input data that follow the same distribution as those in $\mathcal{D}$.

For our \gls{MINN} architecture, we use the channel agnostic \textit{Encoder} and \textit{Decoder} modules $\bar{\myvec{s}}(t) = f^{\rm e}_{\myvec{w}_{\rm e}}(\myvec{x}(t))$ and $\hat{\myvec{o}}(t) = f^{\rm d}_{\myvec{w}_{\rm d}}(\mathbf{y}(t))$ from ~\eqref{eq:encoder-channel-agnostc} and~\eqref{eq:decoder-channel-agnostc}, respectively, as before.
We incorporate a new, \emph{Power Control} module, $P{(t)} = f^{\rm p}_{\mathbf{w}_{\rm p}}(\mathbf{p}{(t)})$, which receives the current position of the \gls{RX}, $\mathbf{p}{(t)}=[x{(t)},y{(t)}]$, and outputs the power value to be used for $\mathbf{x}{(i)}$'s transmission. Note that the input data are not used by this module allowing it to focus on beneficial power control strategies based solely on $\mathbf{p}{(t)}$'s.

The \emph{Channel} module using a \gls{SIM}, $\mathbf{y}{(t)}=f^{\rm c}_{\bar{\boldsymbol{\omega}}}(\mathbf{x}{(t)}, P{(t)}, \boldsymbol{\mathcal{H}}{(t)})$, implements the \gls{OTA} propagation as in~\eqref{eq:received-signal-RIS_1} with $\bar{\myvec{s}}(t) = P{(t)} \myvec{s}(t)$ incorporating the dynamic power control.
Recall that the channel is parametrized by the vector $\bar{\boldsymbol{\omega}} = [\ddot{\boldsymbol{\omega}}_1^\top, \dots, \ddot{\boldsymbol{\omega}}_M^\top]^\top \! \in \! [0,2\pi]^{N}$ containing all \gls{SIM} elements' tunable responses, 
which are treated equivalently to trainable \gls{DNN} parameters. 
Aggregating the parameters of all subsequent modules into the vector $\mathbf{\bar{w}}\!\triangleq\![\mathbf{w}_{\rm t}^\top, \mathbf{w}_{\rm p}^\top, \bar{\boldsymbol{\omega}}^\top,\mathbf{w}_{\rm r}^\top]^\top$ results in the proposed \gls{MINN}:
\begin{equation}
    \mathbf{\hat{o}}(t) = f^{\rm r}_{\mathbf{w}_{\rm r}}(f^{\rm c}_{\bar{\boldsymbol{\omega}}}(f^{\rm t}_{\mathbf{w}_{\rm t}}(\mathbf{x}(t)), f^{\rm p}_{\mathbf{w}_{\rm p}}(\mathbf{p}(t)), \boldsymbol{\mathcal{H}}(t))),
\end{equation}
which is optimized to address a relaxed variation of problem~\eqref{eq:problem-power-control}. Specifically, by integrating the constraint~\eqref{eq:power-constrint} into the objective of~\eqref{eq:objective}, we obtain the following design formulation for $\mathbf{\bar{w}}$:
\begin{subequations}
\begin{align}\label{eq:relaxed-objective}
    \min_{\mathbf{\bar{w}}}\mathcal{\hat{L}}(\mathbf{\bar{w}}) &\triangleq \mathbb{E}_{\boldsymbol{\mathcal{H}}} \left[ \sum_{i=1}^{\mathcal{|D|}}\mathfrak{J}_{\rm CE}\left(\mathbf{\hat{o}}(i), \mathbf{o}(i)\right) + \gamma P(i) \right] \\
    {\rm s.t.}~~& \{ \mathbf{\hat{o}}{(i)}, P{(i)} \} = f_{{\bar{\mathbf{w}}}}(\mathbf{x}{(i)}, \boldsymbol{\mathcal{H}}{(i)}),\,\, i=1,\dots,|\mathcal{D}|, \\
    & [\bar{\boldsymbol{\omega}}_m]_n \in [0, 2\pi],\,\, \forall n=1,\dots, N_m, \, m=1,\dots,M, \label{eq:theta-constraint}
\end{align}
\end{subequations}
where $\gamma$ represents a Lagrange multiplier, empirically tuned to satisfy the $P_{\rm max}$ constraint.
To ensure that all phase shifts reside in the $[0,2\pi]$ range (satisfying~\eqref{eq:theta-constraint}), the operation $\bar{\boldsymbol{\omega}} \gets \pi({\rm tan}^{-1}(\bar{\boldsymbol{\omega}})+1)$ is applied to them.
Throughout the remainder, $(\mathbf{x}(t),\mathbf{o}(t))$ and $\boldsymbol{\mathcal{H}}(t)$ are assumed to be statistically independent, implying the data has no correlation with the wireless environment.
This assumption guarantees that independently sampled random pairs of $(\mathbf{x}(t),\mathbf{o}(t))$ and $\boldsymbol{\mathcal{H}}(t)$ yield unbiased estimators for the expectation of $\mathcal{\hat{L}}(\mathbf{\bar{w}})$~\cite{Robbins_Monro}.
Consequently, we employ \gls{SGD} with backpropagation~\cite{Rumelhart86} on~\eqref{eq:relaxed-objective} to determine approximately optimal values for $\mathbf{\bar{w}}$.
The chain rule derivation for the partial derivatives of the \gls{DNN} module weights follows the derivations of Section~\ref{subsec:Training}.
Crucially, the digital \gls{DNN} modules do not require channel matrix information during the forward pass (i.e., during deployment), as $f^{\rm c}_{\boldsymbol{\omega}}(\cdot)$ is implemented \gls{OTA}.
However, training necessitates an analytical model for $f^{\rm c}_{\boldsymbol{\omega}}(\cdot)$ and knowledge of $\boldsymbol{\mathcal{H}}(t)$ to facilitate gradient backpropagation to the \textit{Encoder} and \textit{Power Control} modules.

\begin{figure}[t]
    \centering
    \includegraphics[width=0.9\linewidth]{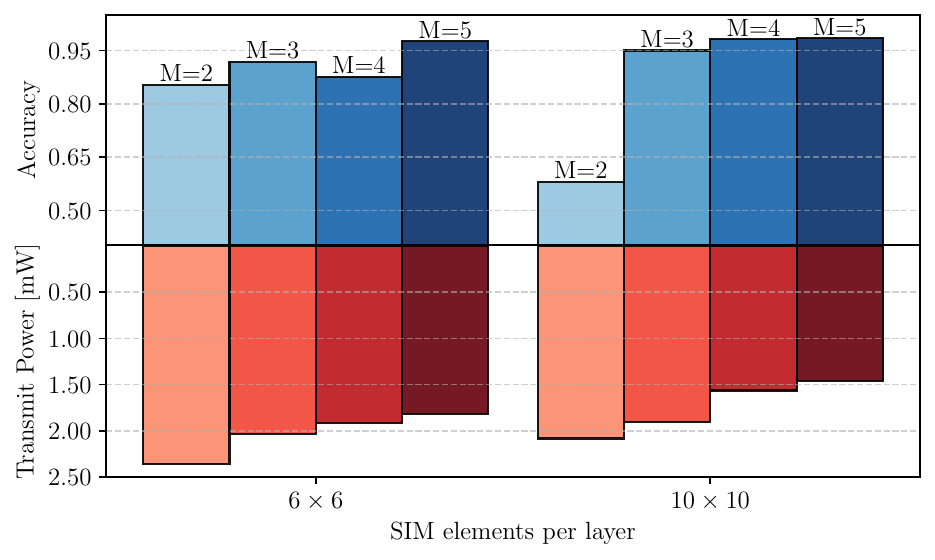}
    \caption{Achieved MNIST classification accuracy and corresponding transmission power during inference for different numbers of \gls{SIM} layers and different numbers of elements per layer considering \gls{MINN} with integrated power control. The accuracy-power tradeoff parameter $\gamma$ is set to $10^{-2}$ for this case.}
    \label{fig:acc-power-barplot}
\end{figure}

\begin{figure}[t]
    \centering
    \includegraphics[width=0.9\linewidth]{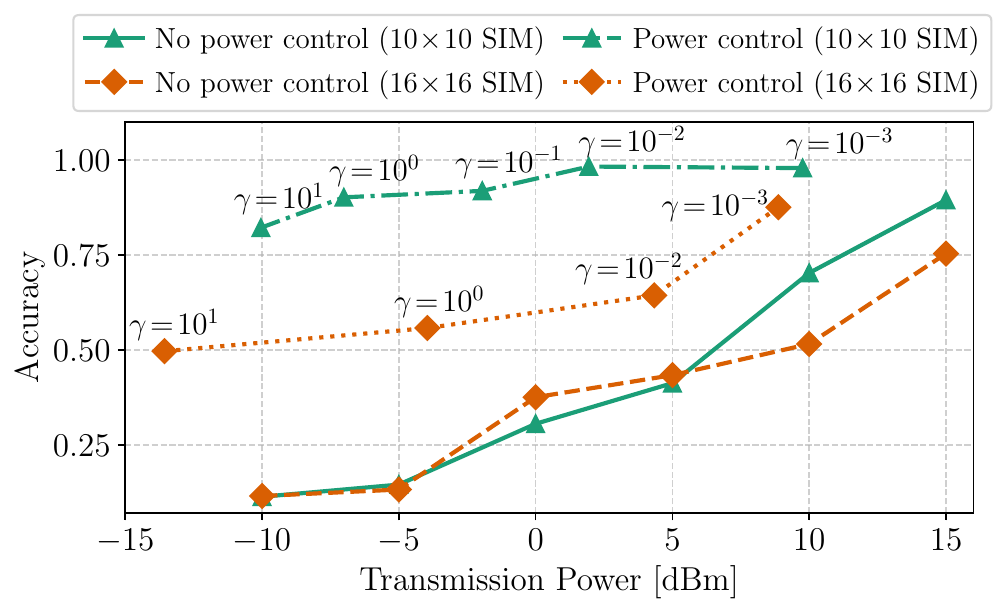}
    \caption{Achieved MNIST classification accuracy for \glspl{MINN} of $4$ \gls{SIM} layers with integrated power control with different tradeoff $\gamma$ values compared to \glspl{MINN} trained with fixed transmission power levels.}
    \label{fig:acc-power-plot-different-gamma}
\end{figure}

To numerically evaluate the performance of the \gls{MINN} with integrated power control we continue the MNIST classification scenarios, considering $16 \times 8$ \gls{MIMO} channels from the Saleh-Valenzuela model~\cite{SV_Model} with $20$ fixed scatterers, while we randomly place the \gls{RX} within the wireless environment at every time step to create dynamic fading.
We allow the training to take place over $200$ epochs under the following procedure:
During the first $30$ epochs, the \textit{Power Control} module $f^{\rm p}_{\mathbf{w}_{\rm p}}(\mathbf{p}{(t)})$ does not partake in the process. Instead $P(t)$ is set to $30$ dBm and both the forward and backward passes involve the \textit{Encoder}, \textit{Decoder}, and \textit{Channel} modules, following the typical \gls{MINN} description presented in the previous sections. After the \nth{30} epoch, $f^{\rm p}_{\mathbf{w}_{\rm p}}(\mathbf{p}{(t)})$ is inserted into the training loop and $\mathbf{w}_{\rm p}$ is updated as part of $\bar{\mathbf{w}}$.
The motivation for this is that \glspl{DNN} perform most of their learning during the first few iterations, where the objective function decreases rapidly.
It is therefore especially important for the highest quality of data to be present at this earlier stage.
In fact, introducing the \textit{Power Control} module's trainable parameters at a later stage during training can be seen as a form of transfer learning and domain adaptation~\cite{TL_tutorial}, where the network learns to adapt to different conditions, imposed by the penalty term of the objective function.

Considering different numbers of layers and elements per layers, classification accuracy results are depicted in Fig.~\ref{fig:acc-power-barplot}, where each case is accompanied by the minimum power attained by the \gls{MINN} while achieving the reported accuracy. It can be inferred that deeper architectures provide improvements in terms of accuracy, but more crucially, in terms of energy efficiency.
Moreover, the effects of the Lagrange multiplier are investigated in Fig.~\ref{fig:acc-power-plot-different-gamma}.
In the figure, \glspl{MINN} with integrated power control are compared against counterparts trained with fixed transmission powers during training to illustrate the benefits of dynamic power manipulation.
As it can be seen, under the same power budgets, \glspl{MINN} with integrated power control achieve considerably larger accuracy results, especially under very low \gls{SNR}.
In fact, for the case of the $10\times10$ \gls{SIM}, the \gls{MINN} without power control achieves comparable accuracy only at the cost of $25$ dB higher power consumption.

\section{Open Challenges}

Despite the growing maturity of \gls{MS} technology~\cite{BAL24_RIS_review, Tsinghua_RIS_Tutorial_ALL}, \gls{D2NN} prototypes, and relevant algorithmic approaches, \glspl{MINN} remain in their infancy.
The primary limitations of this technology highlight several critical avenues for future research, as outlined below.

\textbf{Nonlinearity:}
Existing \gls{MS} designs and models rely principally on linear computations.
Consequently, the \textit{Channel}module acts as a {\em single linear} layer, regardless of how many \gls{SIM} layers are utilized.
These basic linear layers offer only limited approximation capabilities.
To realize truly deep \gls{OTA} architectures, nonlinear activation functions must be integrated into the layer-to-layer metasurface propagation.
This could be achieved through advanced metamaterial designs capable of exhibiting nonlinear responses in the \gls{RF} domain.
Emerging metamaterials or passive/near-passive \gls{RF} circuits offer promising prospects in this regard.
Notably, these nonlinear responses do not necessarily require controllability, as activation functions in \glspl{DNN} are typically fixed~\cite{Stylianopoulos_MIMO_ELM}.

\textbf{Advanced \gls{DNN} architectures:}
Beyond achieving greater depth, state-of-the-art models utilize sophisticated layer architectures that surpass the capabilities of the fully-connected feedforward propagation found in \glspl{D2NN}.
While recent efforts have successfully implemented convolutional layers by leveraging wideband characteristics~\cite{AirCNN}, the implementation of deep convolutional neural networks, or even more complex recurrent and attention-based architectures remains, a formidable challenge.

\textbf{Theoretical guarantees:}
Regardless of the specific \gls{MS} design or layer architecture selected, establishing the universal approximation properties when accounting for the operations induced by the channel is a difficult task.
For \glspl{SLFN} where channel fading coefficients can play the role of random weight parameters in the hidden layer, it can be shown~\cite{Stylianopoulos_MIMO_ELM} that this simplified \gls{MINN} architecture exhibits the characteristics of universal approximation.
This theoretical result holds by casting the \gls{MINN} under the \gls{ELM} framework under the following conditions: (i) static Rayleigh fading, (ii) arbitrarily large, yet finite, $N_r$, and (iii) nonlinear activation function at the \gls{RX}, which may be implemented using analog \gls{RF} components.
Further theoretical advancements are required to extend the guarantees associated with random weights to deeper structures and dynamic fading environments.
Complementary analytical insights could also guide best practices for regularization, initialization, and hyper-parameter selection, leading to more stable training behavior.

\textbf{Exploitation of temporal and frequency degrees of freedom:}
To date, D\textsuperscript{2}NNs and \glspl{MINN} have primarily exploited spatial degrees of freedom via multiplexing across multiple antenna/\gls{MS} elements.
However, temporal and frequency degrees of freedom remain underexplored~\cite{Space_time_coding}.
By leveraging wideband and frequency-selective \gls{MS} responses~\cite{Luo2019_Wideband_D2QN}, the duality between time and frequency domains can be effectively exploited.
In this context, reconfigurable \glspl{MS} could enable feature extraction across multiple transmissions.
Furthermore, equipping the system with temporal memory beyond linear time-invariant limits offers a potential pathway for implementing recurrent layers.

\textbf{OTA training:}
Training for \gls{MINN} and \glspl{D2NN} is typically assumed to occur in simulation, where gradient updates are computed digitally before the resulting responses are forwarded to the \glspl{MS}.
However, more advanced \gls{MS} designs may eventually allow for the objective function and corresponding gradients to be calculated \gls{OTA}.
This would enable the backward pass to be performed in the wave domain by reconfiguring \gls{MS} responses according to impinging error signals.

\section{Conclusions}
\gls{GOC} are gaining momentum in the emerging data-driven networks as they reduce the amount of data transmitted and the associated computations.
Performing \gls{EI}, in particular, involves \gls{DNN}-based \textit{Encoder} and \textit{Decoder} modules at the \gls{TX} and \gls{RX} devices, respectively.
To this end, the whole system can be treated as a single \gls{DNN} and be trained through backpropagation over the channel that induces fading effects.
By employing various types of \glspl{MS} onto the wireless environment environment, the propagation of \gls{RF} signals can be controlled to perform operations similar to linear layers of \gls{DNN} \gls{OTA}.
The basic principle relies on \gls{SIM}-based computational units that implement \glspl{D2NN} for wave-domain \gls{ML}.
\gls{SIM} or \gls{RIS} devices can further be positioned inside the wireless channel so that they shape the resulting dynamic fading to perform desirable \gls{ML} computations.
Consequently, the \textit{Encoder}, \gls{MS}-parametrized \textit{Channel}, and \textit{Decoder} modules now comprise an \gls{E2E} trainable \gls{MINN} architecture with \gls{OTA} computation capabilities that offload part of the processing onto the channel, therefore reducing the overall computational power consumption.
Different variations may be supported, including channel-aware or channel-agnostic transceivers and \glspl{MS} with fixed or dynamically reconfigurable responses, that achieve performance on par with their fully digital counterparts.
By further incorporating dynamic power control during training, \gls{MINN} architectures may learn to reduce the transmission power needed during inference.
Once important challenges related to nonlinear \gls{MS} responses and advanced \gls{ML} operations are overcome, \gls{MINN} are foreseen to constitute a key pillar of energy-efficient intelligent designs for autonomous network applications under communications and sensing.

\FloatBarrier
\bibliographystyle{IEEEtran}
\bibliography{IEEEabrv, refs}

\end{document}